\documentclass[fleqn,usenatbib]{mnras}

\usepackage{newtxtext,newtxmath}

\usepackage[T1]{fontenc}

\DeclareRobustCommand{\VAN}[3]{#2}
\let\VANthebibliography\thebibliography
\def\thebibliography{\DeclareRobustCommand{\VAN}[3]{##3}\VANthebibliography}

\usepackage{graphicx}	
\usepackage{amsmath}	
\usepackage{physics}
\usepackage{bm}
\usepackage{comment}
\usepackage{pdflscape}
\usepackage[dvipsnames]{xcolor}
\usepackage{xspace}

\newcommand{\Msun}{M_{\sun}}
\newcommand{\hMsun}{h^{-1} \, M_{\sun}}
\newcommand{\Mpc}{\mathrm{Mpc}}

\newcommand{\hMpc}{h^{-1} \, \mathrm{Mpc}}

\newcommand{\HA}{H$\alpha$\xspace}
\newcommand{\OII}{[\ion{O}{ii}]\xspace}

\title[Clustering of ELGs with IllustrisTNG I]
{Clustering of emission line galaxies with IllustrisTNG I.:
fundamental properties and halo occupation distribution}

\author[K. Osato and T. Okumura]{
Ken Osato$^{1,2,3}$\thanks{E-mail: ken.osato@chiba-u.jp}
and Teppei Okumura$^{4,5}$
\\
$^{1}$Center for Frontier Science, Chiba University,
1-33 Yayoi-cho, Inage-ku, Chiba 263-8522, Japan\\
$^{2}$Department of Physics, Graduate School of Science, Chiba University,
1-33 Yayoi-cho, Inage-ku, Chiba 263-8522, Japan\\
$^{3}$Center for Gravitational Physics and Quantum Information,
Yukawa Institute for Theoretical Physics, Kyoto University,\\
Kitashirakawa Oiwakecho, Sakyo-ku, Kyoto 606-8502, Japan\\
$^{4}$Academia Sinica Institute of Astronomy and Astrophysics,
No. 1, Section 4, Roosevelt Road, Taipei 10617, Taiwan\\
$^{5}$Kavli Institute for the Physics and Mathematics of the Universe,
The University of Tokyo Institutes for Advanced Study,\\
5-1-5 Kashiwanoha, Kashiwa-shi, Chiba 277-8583, Japan
}

\date{Accepted XXX. Received YYY; in original form ZZZ
\\
Report number: YITP-22-61}

\pubyear{2022}

\begin{document}
\label{firstpage}
\pagerange{\pageref{firstpage}--\pageref{lastpage}}
\maketitle

\begin{abstract}
Upcoming spectroscopic redshift surveys use emission line galaxies (ELGs)
to trace the three-dimensional matter distributions
with wider area coverage in the deeper Universe.
Since the halos hosting ELGs are young and undergo infall
towards more massive halos along filamentary structures,
contrary to a widely employed luminous red galaxy sample,
the dynamics specific to ELGs should be taken into account
to refine the theoretical modelling at non-linear scales.
In this paper, we scrutinise the halo occupation distribution (HOD)
and clustering properties of ELGs by utilising
IllustrisTNG galaxy formation hydrodynamical simulations.
Leveraging stellar population synthesis technique
coupled with the photo-ionization model,
we compute line intensities of simulated galaxies
and construct mock \HA and \OII ELG catalogues.
The line luminosity functions and
the relation between the star formation rate
and line intensity are well consistent with observational estimates.
Next, we measure the HOD and demonstrate that
there is a distinct population for the central HOD,
which corresponds to low-mass infalling halos.
We then perform the statistical inference of HOD parameters
from the projected correlation function.
Our analysis indicates that the inferred HODs significantly deviate
from the HOD measured directly from simulations
although the best-fit model yields a good fit to the projected correlation function.
It implies that the information content of the projected correlation function
is not adequate to constrain HOD models correctly and thus,
it is important to employ mock ELG catalogues
to calibrate the functional form of HOD models
and add prior information on HOD parameters to robustly determine the HOD.
\end{abstract}




\section{Introduction}
\label{sec:introduction}
Matter fluctuations generated during the inflationary era are amplified
through gravitational instability driven primarily by dark matter,
resulting in hierarchical structures which range from stars to galaxy clusters
\citep{White1991}.
Thus, the nature of dark matter and gravitational physics is imprinted onto
the formation and evolution of
the large-scale structures (LSS) of the Universe
\citep[for a review, see][]{Weinberg2013}.
In addition, dark energy induces accelerated expansion of the Universe
and has considerable impacts on the geometry of the Universe.
The geometry of the Universe is robustly measured from
the acoustic scale of sound waves in the cosmic baryon-photon fluid,
referred to as baryon acoustic oscillation (BAO).
The BAO scale serves as the standard ruler of the Universe,
and the measurement of the scale can tightly constrain
cosmological parameters \citep{Eisenstein2005,Aubourg2015}.
Though the fundamental information of the Universe is embedded into
the matter distribution,
it is dominated by dark matter invisible to optical telescopes.
Instead, one can use galaxies as surrogates for the matter distribution
because they generally form in dense regions and thus
are expected to trace the matter distribution.
To do this, galaxy clustering analysis, which has been playing
a central role in observational cosmology, is commonly exploited.
However, since galaxy formation processes are governed by complex nonlinear
astrophysics, the relation between galaxy and matter distributions becomes nonlinear,
which makes it challenging to study it from the first principle.
This \textit{galaxy bias} \citep[for a review, see][]{Desjacques2018}
is considered as a critical obstacle for galaxy clustering analysis to date.

In the last decade, various observational programmes, which aim to
measure galaxy clustering and constrain cosmological models from it,
have been conducted:
extended Baryon Oscillation Spectroscopic Survey \citep[eBOSS;][]{Alam2021},
6dF Galaxy Survey \citep{Beutler2011},
WiggleZ \citep{Blake2011}, and
VIPERS \citep{delaTorre2017}.
Recently, wider and deeper spectroscopic surveys are being proposed to improve the accuracy:
Subaru Prime Focus Spectrograph \citep[PFS;][]{Takada2014},
Dark Energy Spectrograph Instrument \citep[DESI;][]{DESI2016a,DESI2016b},
\textit{Euclid} \citep{Laureijs2011,Amendola2013}, and
Roman Space Telescope\footnote{\url{https://roman.gsfc.nasa.gov}}.
These upcoming spectroscopic surveys will target a specific population of galaxies:
emission line galaxies (ELGs).
ELGs are characterized by the bright line emissions, e.g.,
\HA ($\lambda \, 6564.62 \, \mathrm{\mathring{A}}$) or
\OII ($\lambda \lambda \, 3727.09 \, \mathrm{\mathring{A}}, 3729.88 \, \mathrm{\mathring{A}}$),
which are sourced from circumstellar gas irradiated by massive OB-type stars.
The lifetime of such massive stars is short and
the nebular emission lasts only for $\simeq 10 \, \mathrm{Myr}$ \citep{Byler2017}.
In summary, the emission lines are an indicator of star-formation activity
\citep{Charlot2001,Kewley2019}.
The epoch at the redshift $z = 1 \text{--} 2$ is called \textit{cosmic noon},
where star formation is the most active over the cosmic history,
and at this epoch, a large number of ELGs are expected to be detected.
Several pilot surveys have measured ELG clustering
around cosmic noon, e.g.,
HiZELS \citep{Geach2008,Sobral2010} and
FastSound \citep{Tonegawa2015,Okada2016,Okumura2016}.
In the near future, large survey programmes will improve
the number of detected ELGs and survey areas,
and thus, the widest galaxy clustering analysis
at the deepest Universe will be accessible.

To extract maximal information from the observed galaxy distribution,
it is critical to develop accurate and precise theoretical models 
to predict statistics of galaxy clustering down to small (non-linear) scales.
Spectroscopic redshift surveys used to
target mainly stellar-mass limited samples,
which are widely referred to as luminous red galaxies \citep[LRGs;][]{Eisenstein2001}.
In general, LRGs are red and quenched galaxies and
thus are likely to be hosted by old and massive halos.
So far, various theoretical models have been developed to predict LRG clustering.
For example, the abundance matching technique \citep{Vale2004,Conroy2006},
which connects the luminosity of LRGs to
halo properties, e.g., halo mass or circular velocity,
can reproduce the observed correlation functions quite well.
However, it is not trivial whether the same theoretical frameworks
can be applicable to ELG clustering.
In fact, previous works \citep{Orsi2018,GonzalezPerez2018,GonzalezPerez2020}
used semi-analytic simulations,
which populate galaxies onto dark-matter-only $N$-body simulations
based on the merger history of dark halos
\citep{Springel2005,Croton2016}, highlighted
the difference of LRG and ELG clustering;
LRGs are well virialized in halos and located close to the halo centres
but a majority of halos hosting ELGs is undergoing infall towards massive halos
along filamentary structures.
The nebular emission of ELGs lasts only for $10 \, \mathrm{Myr}$, which corresponds to the lifetime of the central star,
but the dynamical time scale of the hosting halo is $\gtrsim 100 \, \mathrm{Myr}$.
Therefore, it is expeted that a large fraction of the nebular emission is extinguished during the first passage to the pericentre.
Furthermore, it is shown that
the velocity distribution of ELGs within a host halo significantly
deviates from Gaussian distribution due to the infall motion \citep{Orsi2018}.
This coherent dynamics specific to ELGs
has appreciable impacts on theoretical modelling of the velocity field,
i.e., ELG clustering in redshift space.

In order to shed light on the ELG clustering,
galaxy formation hydrodynamical simulations
are one of the most promising and powerful methods
because the simulations can trace the evolution of LSS
and the formation of ELGs simultaneously.
The galaxy formation physics involves
formation and evolution of various astronomical objects
with the wide dynamic range; from $\mathrm{AU}$ scale for black holes
to $\mathrm{Mpc}$ scale for galaxy clusters.
Such multi-scale physics can be approximated as subgrid physics,
which implements galaxy formation physics by coarse-graining
in an ad-hoc manner \citep[for a review, see][]{Somerville2015}.
The subgrid hydrodynamical simulations can
better reproduce observed quantities, such as
stellar mass functions and black hole co-evolution,
once free parameters are properly calibrated.
In this paper, we exploit the hydrodynamical simulations
to generate high-fidelity mock ELG catalogues
and address clustering of the simulated ELGs for better understanding of
clustering nature of ELGs.
Compared to the semi-analytical approach,
the key advantage of hydrodynamical simulations is that
a wide variety of information, e.g., formation age, environments, and dynamics,
are readily available and enable us to create more realistic ELG catalogues
\citep{Park2015,Shimizu2016,Favole2020,Hadzhiyska2021,Favole2022,Yuan2022}.
Therefore, it is possible to scrutinise the relation between
the observed properties of ELGs and large-scale clustering,
which reflects the physics of dark matter and dark energy.
We construct mock catalogues of \HA and \OII ELGs, which are targets
of \textit{Euclid} and PFS surveys, respectively, by leveraging
the state-of-the-art hydrodynamical simulations, IllustrisTNG, 
coupled with the stellar population synthesis model.
Then, we measure halo occupation distribution (HOD), i.e.,
the mean number of galaxies as the function of host halo mass,
which is the straightforward way to relate ELGs to host halos.
Finally, we measure the projected correlation function,
which is one of major statistics used in actual observations,
and carry out an \textit{HOD challenge},
where we investigate how one can reproduce the underlying HOD
from statistical analysis of the measured projected correlation function.
In subsequent series of papers, we will present
a cosmological parameter challenge analysis with correlation functions in redshift space
and detailed analysis on ELG-halo connection, in particular assembly bias and velocity bias.

This paper is organized as follows.
In Section~\ref{sec:simulations}, we present numerical methods
to generate mock ELG catalogues used in this work:
IllustrisTNG hydrodynamical simulations and post-processing
with stellar population synthesis code \texttt{P\'EGASE-3}.
In Section~\ref{sec:theory}, we overview HOD
and theoretical modelling of projected correlation functions.
In Section~\ref{sec:fundamental_properties}, we discuss the properties of mock ELGs
and present the measurement of HODs of mock ELGs.
In Section~\ref{sec:inference_HOD}, we constrain
HOD model parameters from the direct measurement of the HOD
and the projected correlation function.
We make a concluding remark in Section~\ref{sec:conclusions}.

\section{Numerical Simulations}
\label{sec:simulations}
In this section, we overview our numerical methods to produce the mock catalogues of
\HA and \OII ELGs. 

\subsection{Hydrodynamical simulations}
As the backbone of our mock simulation suite,
we utilise IllustrisTNG simulations
\citep{Nelson2018,Pillepich2018b,Springel2018,Naiman2018,Marinacci2018,Nelson2019}.
The simulations are run by the moving mesh code \texttt{AREPO} \citep{Springel2010}
and various baryonic processes are implemented as subgrid physics \citep{Weinberger2017,Pillepich2018a}.
Among the realisations of the simulation suite,
we use the run with the largest box, TNG300-1 (hereafter TNG300),
where the length on a side is $205 \, \hMpc$ in comoving coordinates,
the number of dark matter particles and gas cells is $2500^3$ for each,
and the corresponding mass resolution for dark matter and gas is
$m_\mathrm{DM} = 5.9 \times 10^7 \, h^{-1} \, \Msun$ and
$m_\mathrm{gas} = 1.4 \times 10^6 \, h^{-1} \, \Msun$, respectively.
The simulation snapshots are densely sampled
in the wide range of redshifts $z = 0 \text{--} 20$.
Unless stated, we use the snapshot at the redshift $z = 1.5$,
which is the typical redshift for future spectroscopic surveys.

Throughout this paper, a flat $\Lambda$-cold dark matter Universe is assumed
and cosmological parameters are best-fit parameters of the \textit{Planck} 2015 results
\citep{Planck2015XIII}, as used in the IllustrisTNG simulations;
Hubble constant $H_0 = 67.74 \, \mathrm{km} \, \mathrm{s}^{-1} \, \Mpc^{-1}$,
the baryon density $\Omega_\mathrm{b} = 0.0486$,
the total matter density $\Omega_\mathrm{m} = 0.3089$,
the spectral index of scalar perturbations $n_\mathrm{s} = 0.9667$,
and the amplitude of matter fluctuations at $8\,\hMpc$ $\sigma_8 = 0.8159$.
For neutrinos, we assume massless neutrinos
with an effective number of neutrino species $N_\mathrm{eff} = 3.046$.

\subsection{Stellar population synthesis}
\label{sec:SPS}
Since the emission line intensities are not provided in the IllustrisTNG snapshots,
we post-process the output to calculate
spectral energy distributions (SEDs) of simulated galaxies
based on stellar population synthesis (SPS).
For this purpose, we employ the SPS code \texttt{P\'EGASE-3} \citep{Fioc2019},
which calculates SEDs from far-IR to UV wavelengths and
also computes the nebular emission intensities based on
the photo-ionization code \texttt{CLOUDY} \citep{Ferland2017}.
In \texttt{P\'EGASE-3},
we adopt the initial mass function of \citet{Chabrier2003},
Padova stellar evolutionary tracks \citep{Bressan1993,Fagotto1994a,Fagotto1994b,Fagotto1994c,Gicardi1996},
stellar yields for low mass stars of \citet{Marigo2001} and
high mass stars of \citet{Portinari1998},
and stellar spectra libraries of BaSeL v3.1 \citep{Westera2002}.

First, we run \texttt{P\'EGASE-3}
for eight initial metallicities,
$Z_\mathrm{ini} = (0.0, 0.0001, 0.0004, 0.004, 0.008, 0.02, 0.05, 0.1)$,
where \texttt{CLOUDY} precomputed tables are available,
and then SEDs and line intensities are output every $1 \, \mathrm{Myr}$
in the range of $0 \text{--} 100 \, \mathrm{Myr}$.
Thus, we construct a precomputed table for line intensity with respect to
metallicity and age.
Next, we compute line luminosities for each galaxy found in TNG300.
The subhalo catalogues provide the list of stellar particles which constitute
a subhalo, and we regard a subhalo as a simulated \textit{galaxy}.
The contribution from a single stellar particle is calculated
by bi-linear interpolation with the precomputed table
with respect to the formation age and the metallicity,
both of which are available in particle snapshots.
In a sense, we regard a stellar particle as a tiny isolated galaxy.
Since the line intensity of the precomputed table is normalized by the stellar mass,
we sum up all the particle contributions by multiplying
the initial stellar mass at the formation time\footnote{Note that
the mass of stellar particles gradually decreases due to the stellar evolution.}
and the resultant sum is the line luminosity of the galaxy.

Since the interested emission lines are subject to
absorption due to dusts \citep[for a review, see][]{Salim2020},
the dust attenuation should be taken into account for observed line luminosities.
\citet{Nelson2018} addressed the dust attenuation
from unresolved and resolved dusts
and found that the impact of the resolved dust is relatively small
compared with the unresolved one.
Hence, we consider only unresolved dust contribution for simplicity.
Instead of a power-law extinction model \citep{Charlot2000}
for unresolved dust contribution adopted in previous studies,
in our modelling, the \texttt{P\'EGASE-3} code
computes the dust evolution within a galaxy
and incorporates dust extinction due to dust grains in star-forming clouds
and diffuse inter-stellar medium
\citep[for details, see Section~6.1 of ][]{Fioc2019}.
As discussed in Section~\ref{sec:properties_ELGs},
this prescription gives reasonable estimates of dust attenuation effects.

\section{Theory}
\label{sec:theory}
In this section, we review theoretical models used to analyse
the clustering of ELGs in this paper.
We present the halo model of the projected correlation function
and its covariance matrix in Sections~\ref{sec:correlation_function}
and \ref{sec:covariance}, respectively. 
We then introduce halo occupation models to relate the ELGs to their host halos in Section~\ref{sec:hod}.

\subsection{Projected correlation function}\label{sec:correlation_function}
In order to compute the statistics of spatial clustering of ELGs,
we begin with the galaxy power spectrum.
We adopt a halo model, where all galaxies
are assumed to form inside the host dark halos.
There are two types of galaxies according to the internal locations within a halo:
central and satellite galaxies.
The central galaxies are located at the centre of the hosting dark halos
and in general, the most massive galaxies in the halo.
The satellite galaxies are distributed following the density profile of the halos
and a halo can host multiple satellite galaxies.
Under the halo model prescription,
the galaxy power spectrum has two contributions;
clustering in a single halo denoted as the one-halo term $P_\mathrm{1h}$
and clustering between two distinct halos denoted as the two-halo term $P_\mathrm{2h}$.
The expression of the galaxy power spectrum $P_\mathrm{gg}$ is given as
\begin{align}
  \label{eq:power_galaxy}
  P_\mathrm{gg}  (k, z) =& P_\mathrm{1h} (k, z) + P_\mathrm{2h} (k, z) , \\
  P_\mathrm{1h} (k, z) =& \frac{1}{n_\mathrm{g}^2} \int \!\! \dd{M}
  \dv{n_\mathrm{h}}{M} (M, z)
  [2 \langle N_\mathrm{cen} N_\mathrm{sat} \rangle \tilde{u}_\mathrm{h} (k, M, z)
  \nonumber \\
  & + \langle N_\mathrm{sat} (N_\mathrm{sat} - 1) \rangle \tilde{u}_\mathrm{h}^2 (k, M, z)] , \\
  P_\mathrm{2h} (k, z) =& \frac{1}{n_\mathrm{g}^2} \int \!\! \dd{M_1}
  \dv{n_\mathrm{h}}{M} (M_1, z)
  [\langle N_\mathrm{cen} \rangle + \langle N_\mathrm{sat} \rangle
  \tilde{u}_\mathrm{h} (k, M_1, z) ]
  \nonumber \\
  & \times \int \!\! \dd{M_2} \dv{n_\mathrm{h}}{M} (M_2, z)
  [\langle N_\mathrm{cen} \rangle + \langle N_\mathrm{sat} \rangle
  \tilde{u}_\mathrm{h} (k, M_2, z) ]
  \nonumber \\
  & \times P_\mathrm{hh} (k, M_1, M_2, z) ,
\end{align}
where $\langle N_\mathrm{cen}\rangle$ and $\langle N_\mathrm{cen}\rangle$
are mean occupation numbers of central and satellite galaxies in a halo with mass $M$, respectively,
and $n_\mathrm{g}$ is the galaxy number density.
For the halo mass function, $\dd{n_\mathrm{h}} / \dd{M}$,
we adopt the fitting formula of \citet{Tinker2008}.\footnote{In the fitting formula,
as the halo mass we adopt the mass enclosed by a sphere with the radius within which
the mean density is 200 times the mean density, $M_{200}$.
The mass integration is performed with respect to the virial mass $M_\mathrm{vir}$,
and the halo mass $M_{200}$ is computed by converting the virial mass $M_\mathrm{vir}$
by assuming the NFW profile.}
There are several assumptions made to derive the expressions above.
First, the radial distribution of satellite galaxies follow the density profile
of halos, and we adopt
Navarro--Frenk--White (NFW) profile \citep{Navarro1996,Navarro1997}.\footnote{The assumption that
the radial number density profile of ELGs follows the NFW profile is validated with our mock catalogues
and further investigation of the radial profile will be presented in the subsequent paper of the series.}
In order to specify the NFW profile, $\rho_\mathrm{NFW} (r; M, z)$,
the halo radius and the concentration parameter,
which is defined as the ratio between halo radius and density scale radius,
have to be determined. We adopt the virial radius as the halo radius \citep{Bryan1998}
and the concentration-mass relation of \citet{Klypin2016}
calibrated with $N$-body simulations.
The function $\tilde{u}_\mathrm{h} (k; M, z)$ is the Fourier transform of the scaled density profile,
$\rho_\mathrm{NFW} (r; M, z)/M$, and we use the expression derived analytically by \citet{Scoccimarro2001}.
We assume that the central and satellite occupation numbers are independent, i.e.,
$\langle N_\mathrm{cen} N_\mathrm{sat} \rangle = \langle N_\mathrm{cen} \rangle \langle N_\mathrm{sat} \rangle$,
and the satellite occupation follows the Poisson distribution
$\langle N_\mathrm{sat} (N_\mathrm{sat} - 1) \rangle
= \langle N_\mathrm{sat} \rangle^2$.
Regarding the cross-power spectrum of halos with masses $M_1$ and $M_2$,
$P_\mathrm{hh} (k, M_1, M_2, z)$, higher-order bias contributions can be neglected
at our interested scales and the non-linear effects can be incorporated
by replacing the linear power spectrum with the non-linear one \citep{vandenBosch2013}.
As a result, the cross-power spectrum is approximated as
\begin{equation}
  P_\mathrm{hh} (k, M_1, M_2, z) =
  b_\mathrm{h} (M_1, z) b_\mathrm{h} (M_2, z) P_\mathrm{NL} (k, z) ,
\end{equation}
where $b_\mathrm{h} (M, z)$ is the halo bias, for which we adopt the fitting formula
calibrated with $N$-body simulations \citep{Tinker2010}.
$P_\mathrm{NL} (k, z)$ is the non-linear matter power spectrum
and we adopt the fitting formula
based on the \textit{halofit} prescription \citep{Smith2003}
with parameters calibrated in \citet{Takahashi2012}.

Then, the galaxy three-dimensional correlation function $\xi_\mathrm{gg}$
is computed via a Hankel transform:
\begin{equation}
  \xi_\mathrm{gg} (r, z) = \frac{1}{2\pi^2}
  \int k^2 P_\mathrm{gg} (k, z) \frac{\sin (kr)}{kr} \dd{k} ,
\end{equation}
where the \texttt{FFTLog} algorithm \citep{Hamilton2000} is employed
to compute the Hankel transform.
Finally, the projected correlation function $w_\mathrm{p} (R, z)$ can be computed
by projecting the three-dimensional correlation function along the line-of-sight direction:
\begin{equation}
  w_\mathrm{p} (R, z) = 2 \int_0^{r_\mathrm{max}} \xi_\mathrm{gg}
  \left( r = \sqrt{R^2 + r_\pi^2}, z \right) \dd{r_\pi} ,
\end{equation}
where $r_\mathrm{max}$ is the projection length
and we fix $r_\mathrm{max} = 50 \, \hMpc$ throughout this paper.

\subsection{Covariance matrix of the projected correlation function}
\label{sec:covariance}
In order to perform the statistical analysis with
projected correlation functions, the statistical error estimation,
i.e., covariance matrix, is an essential component.
The most straightforward way to estimate the covariance matrix
is to run many realisations of hydrodynamical simulations
and compute the covariance among the realisations.
However, this approach is not realistic
due to the high computational costs of
hydrodynamical simulations.
There are several options to estimate the covariance matrix
directly from the measurements such as the jackknife method.
However, such methods are not able to estimate the covariance matrix
at large scales.
Therefore, we adopt the analytical method for covariance matrix
calculations to exploit the measurements at all scales.
To compute the covariance matrix of the projected correlation
functions,
we employ the Gaussian covariance matrix \citep{Marian2015}.
The Gaussian covariance matrix is known to
underestimate the covariance
at small scales ($R \lesssim 5 \, \hMpc$)
due to the higher order contribution.
However, such non-linearity can be partially incorporated
by using non-linear galaxy power spectrum in Eq.~\eqref{eq:power_galaxy}.
Thus, we employ the Gaussian covariance matrix
for the statistical analysis.

The analytic expression of the Gaussian covariance matrix
is given as
\begin{align}
  \mathrm{Cov}[w_\mathrm{p}] (R_i, R_j) = \frac{2}{V_\mathrm{s}} \left\{ 2 r_\mathrm{max}
  \int_0^\infty \frac{\dd{k_\perp}}{2 \pi} k_\perp \bar{J}_0 (k_\perp R_i) \bar{J}_0 (k_\perp R_j)
  \right. \nonumber \\
  \left. \times P_\mathrm{gg} (k_\perp) \left[ P_\mathrm{gg} (k_\perp) +
  \frac{2}{n_\mathrm{g}} \right] +
  \frac{\delta_{ij}^\mathrm{K}}{n_\mathrm{g}^2}
  \left[ \bar{w}_\mathrm{p} (R_i) + 2 r_\mathrm{max} \right] \right\} ,
  \label{eq:wp_Cov}
\end{align}
where $\delta_{ij}^\mathrm{K}$ is the Kronecker delta,
$V_\mathrm{s}$ is the survey volume, and
$r_\mathrm{max}$ is the maximum projection length.
In the above equation, we defined the bin-averaged Bessel function, $\bar{J}_0$, and projected correlation function, $\bar{w}_\mathrm{p}$, as
\begin{align}
  \bar{J}_0 (k_\perp R_i) &\equiv \frac{2 \pi}{S_{R_i}}
  \int_{R_{i,\mathrm{min}}}^{R_{i,\mathrm{max}}} J_0 (k_\perp R) R \dd{R} , \\
  \label{eq:bin_averaged_wp}
  \bar{w}_\mathrm{p} (R_i) &\equiv \frac{2 \pi}{S_{R_i}}
  \int_{R_{i,\mathrm{min}}}^{R_{i,\mathrm{max}}} w_\mathrm{p} (R) R \dd{R} ,
\end{align}
where the subscript $i$ denotes the transverse bin,
$R_{i,\mathrm{min}}$ and $R_{i,\mathrm{max}}$ are lower and upper edges
of the $i$-th radial bin, respectively, and
$S_{R_i} = \pi \left( R_{i,\mathrm{max}}^2 - R_{i,\mathrm{min}}^2 \right)$.

\subsection{Parametrisation of halo occupation distributions}
\label{sec:hod}
One of the most fundamental quantities to characterize a dark halo is its mass,
and thus, the occupation number of galaxies
with respect to the halo mass, i.e., halo occupation distribution (HOD),
is commonly used to relate the galaxies to their host halos
\citep{Jing1998,Peacock2000,Seljak2000,Ma2000,Scoccimarro2001,Cooray2002,
Berlind2002,Zheng2005,vandenBosch2013,Hikage2013,Okumura2015}.
In order to describe HOD of LRGs,
\citet{Zheng2005} introduced the following functional form:
\begin{align}
  \langle N_\mathrm{cen} \rangle (M) &= \frac{1}{2} \left[ 1 +
  \mathrm{erf} \left( \frac{\log (M/M_\mathrm{min})}{\sigma_{\log M}} \right) \right] ,
  \\
  \langle N_\mathrm{sat} \rangle (M) &= \frac{1}{2} \left[ 1 +
  \mathrm{erf} \left( \frac{\log (M/M_\mathrm{min})}{\sigma_{\log M}} \right) \right]
  \left( \frac{M-M_0}{M_1} \right)^\alpha ,
\end{align}
where $(M_\mathrm{min}, \sigma_{\log M}, M_0, M_1, \alpha)$ are free parameters.
In this model, the occupation number of central galaxies gradually
reaches unity towards the high mass end
and that of satellite galaxies increases as power law.
This HOD well describes the clustering of
LRG samples \citep{Zehavi2011}.
On the other hand, semi-analytic simulations \citep{Contreras2013,GonzalezPerez2018} and
observations \citep{Geach2012,Avila2020,Okumura2021,Gao2022} indicate
a different shape of HOD for ELGs.
We adopt a modified version of the HOD model of \citet{Geach2012}:
\begin{align}
  \langle N_\mathrm{cen} \rangle (M) =& F_\mathrm{c}
  \exp \left[ - \frac{\log (M/M_\mathrm{c1})^2}{2\sigma_{\log M_\mathrm{c1}}^2} \right]
  \nonumber \\
  & + \frac{1}{2} \left[ 1 +
  \mathrm{erf} \left( \frac{\log (M/M_\mathrm{c2})}{\sigma_{\log M_\mathrm{c2}}} \right) \right] ,
  \label{eq:Geach_cen} \\
  \langle N_\mathrm{sat} \rangle (M) =& F_\mathrm{s} \frac{1}{2} \left[ 1 +
  \mathrm{erf} \left( \frac{\log (M/M_\mathrm{s})}{\sigma_{\log M_\mathrm{s}}} \right) \right]
  \left( \frac{M}{M_\mathrm{s}} \right)^\alpha ,
  \label{eq:Geach_sat}
\end{align}
where $(F_\mathrm{c}, F_\mathrm{s}, M_\mathrm{c1}, \sigma_{\log M_\mathrm{c1}},
M_\mathrm{c2}, \sigma_{\log M_\mathrm{c2}}, M_\mathrm{s}, \sigma_{\log M_\mathrm{s}}, \alpha)$ are free parameters.
In the original HOD of \citet{Geach2012}, similarly to the first term of
Eq.~\eqref{eq:Geach_cen}, the amplitude of the second term
is a free parameter to incorporate the incompleteness.
However, by measuring the HOD directly from simulations,
we found that the completeness of the central galaxies in our samples was close to unity
(see Section~\ref{sec:HOD_measurement}).
We thus fix this amplitude as unity in this paper.
The characteristic point of this HOD model is that
there is a log-normal term (first term of Eq.~\ref{eq:Geach_cen}),
which corresponds to an infalling distinct population.
Hereafter, we refer to this HOD model as the Geach model.

Given an HOD model, one can compute various quantities related to ELGs,
such as the number density,
\begin{equation}
  \label{eq:HOD_ng}
  n_\mathrm{g} = \int \!\! \dd{M} \dv{n_\mathrm{h}}{M} (M, z)
  \left[\langle N_\mathrm{cen} \rangle (M) +
  \langle N_\mathrm{sat} \rangle (M) \right] ,
\end{equation}
the effective bias,
\begin{equation}
  b_\mathrm{eff} = \frac{1}{n_\mathrm{g}} \int \!\! \dd{M}
  \dv{n_\mathrm{h}}{M} (M, z)
  \left[\langle N_\mathrm{cen} \rangle (M) +
  \langle N_\mathrm{sat} \rangle (M) \right] b_\mathrm{h} (M, z),
\end{equation}
the effective host halo mass,
\begin{equation}
  M_\mathrm{eff} = \frac{1}{n_\mathrm{g}} \int \!\! \dd{M}
  \dv{n_\mathrm{h}}{M} (M, z) \left[\langle N_\mathrm{cen} \rangle (M) +
    \langle N_\mathrm{sat} \rangle (M) \right] M,
\end{equation}
and the satellite fraction,
\begin{equation}
  f_\mathrm{sat} = \frac{1}{n_\mathrm{g}} \int \!\! \dd{M}
  \dv{n_\mathrm{h}}{M} (M, z) \langle N_\mathrm{sat} \rangle (M) .
\end{equation}

\section{Fundamental properties of mock ELGs}
\label{sec:fundamental_properties}
In this section, we present properties of our simulated ELGs and compare them
with observational estimates to verify our mock ELG catalogues. We then
measure the HOD of mock ELGs.

\begin{figure}
  \includegraphics[width=\columnwidth]{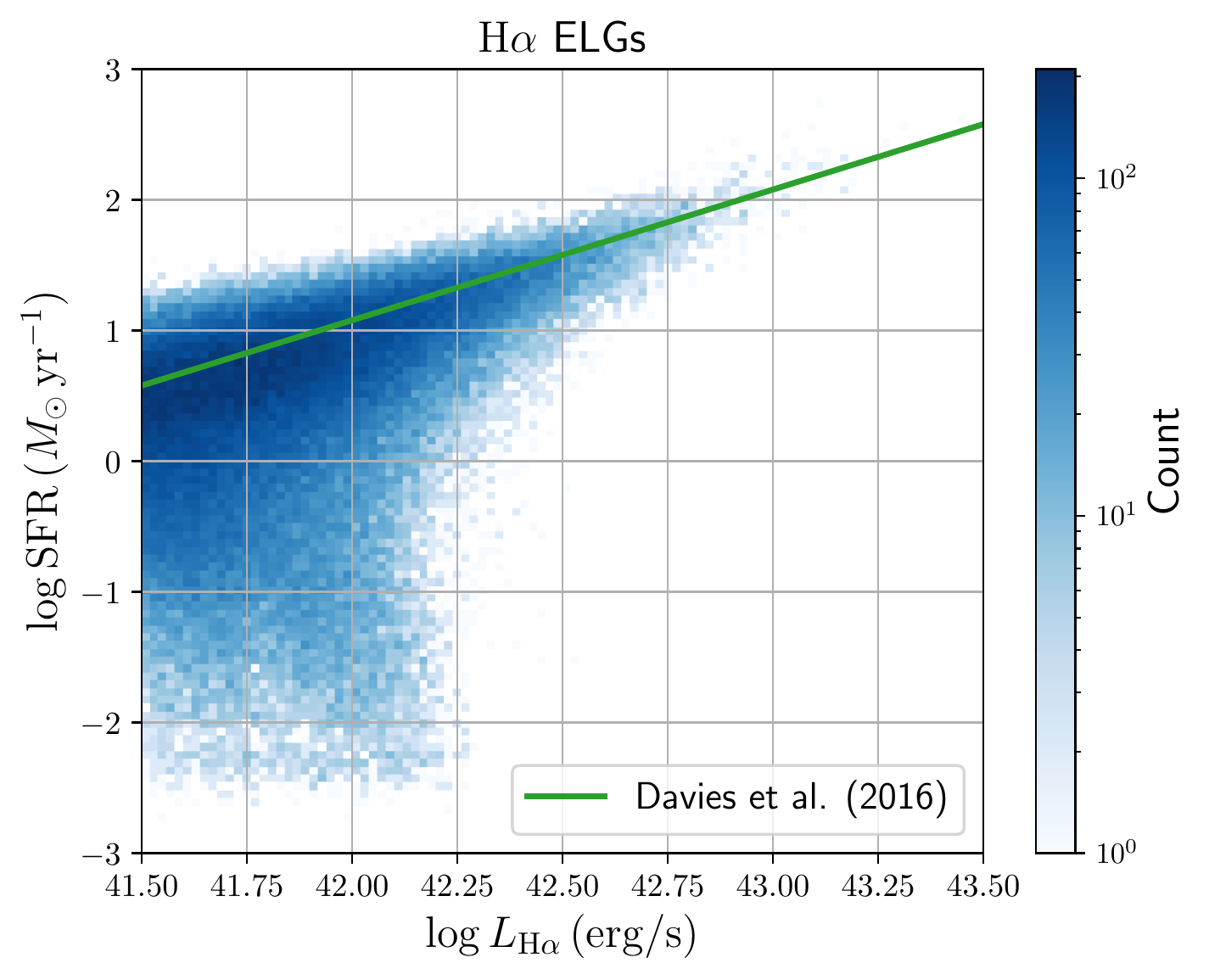}
  \includegraphics[width=\columnwidth]{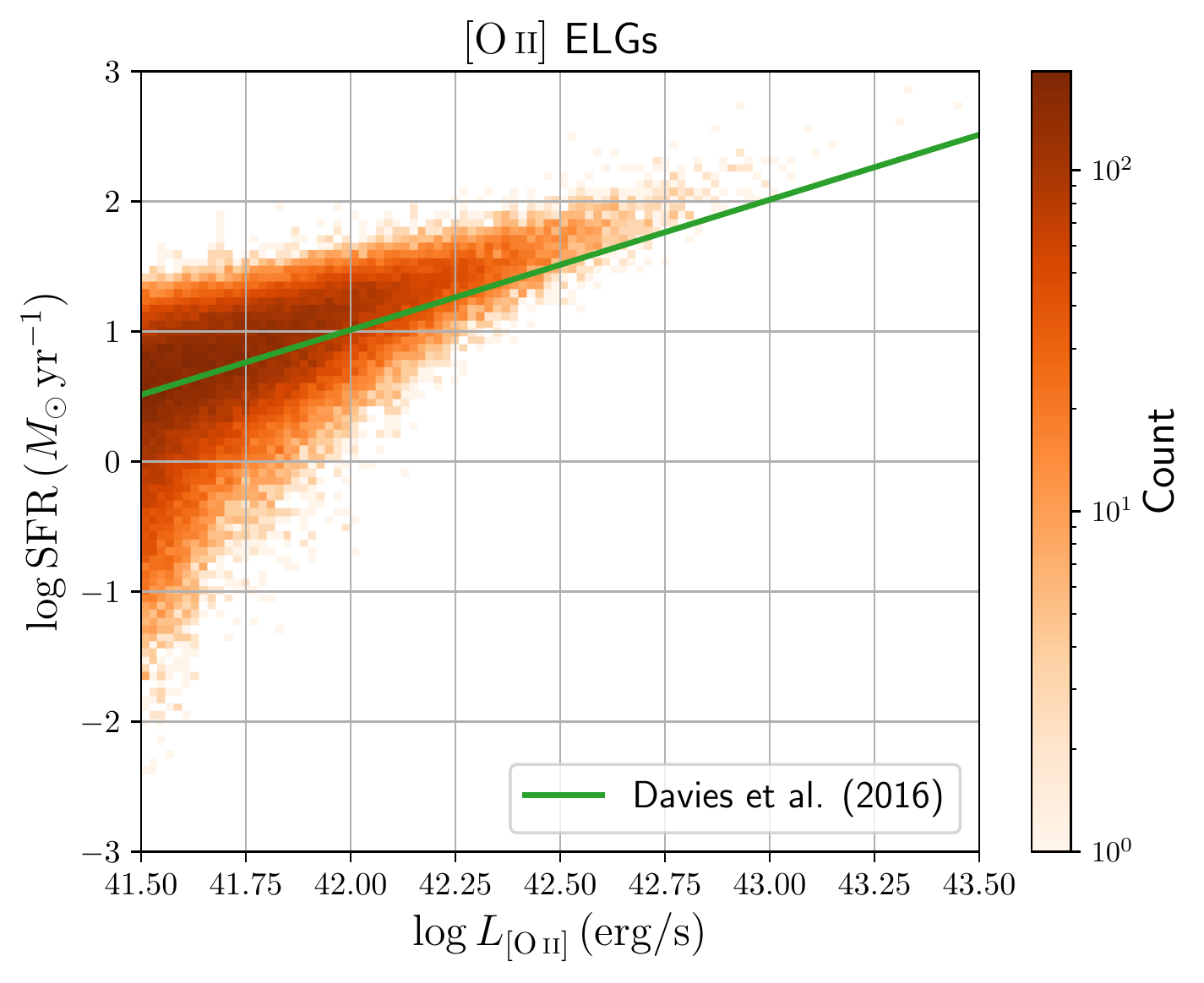}
  \caption{The SFR-line luminosity relation of \HA ELGs (upper panel)
  and \OII ELGs (lower panel).
  For line luminosities, the dust attenuation effect has been considered.
  The power-law relations
  (Eqs.~\ref{eq:SFR_HA} and \ref{eq:SFR_OII})
  based on observational estimates \citep{Davies2016}
  are shown as green solid lines.}
  \label{fig:SFR_Lum}
\end{figure}

\begin{figure}
  \includegraphics[width=\columnwidth]{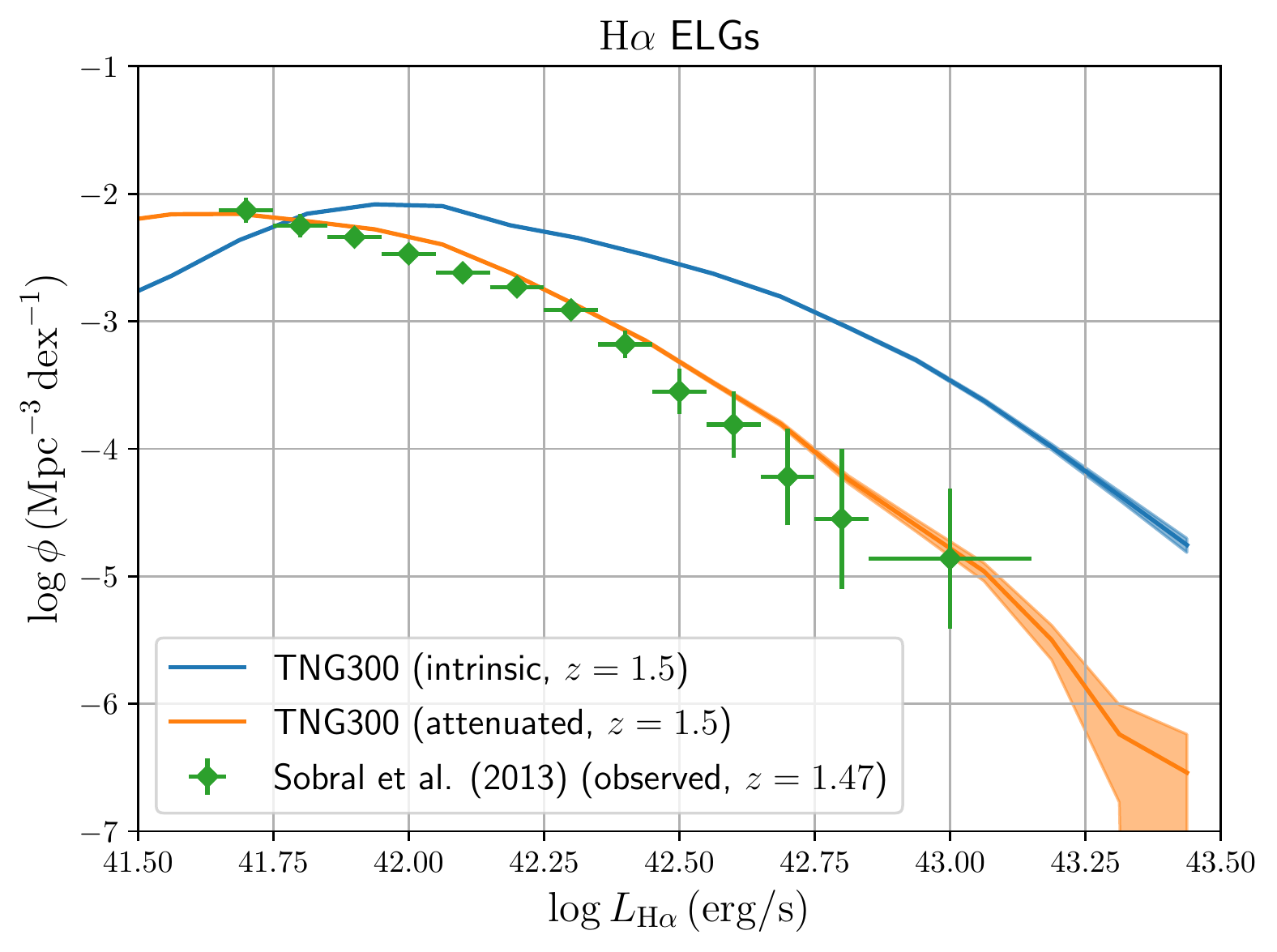}
  \includegraphics[width=\columnwidth]{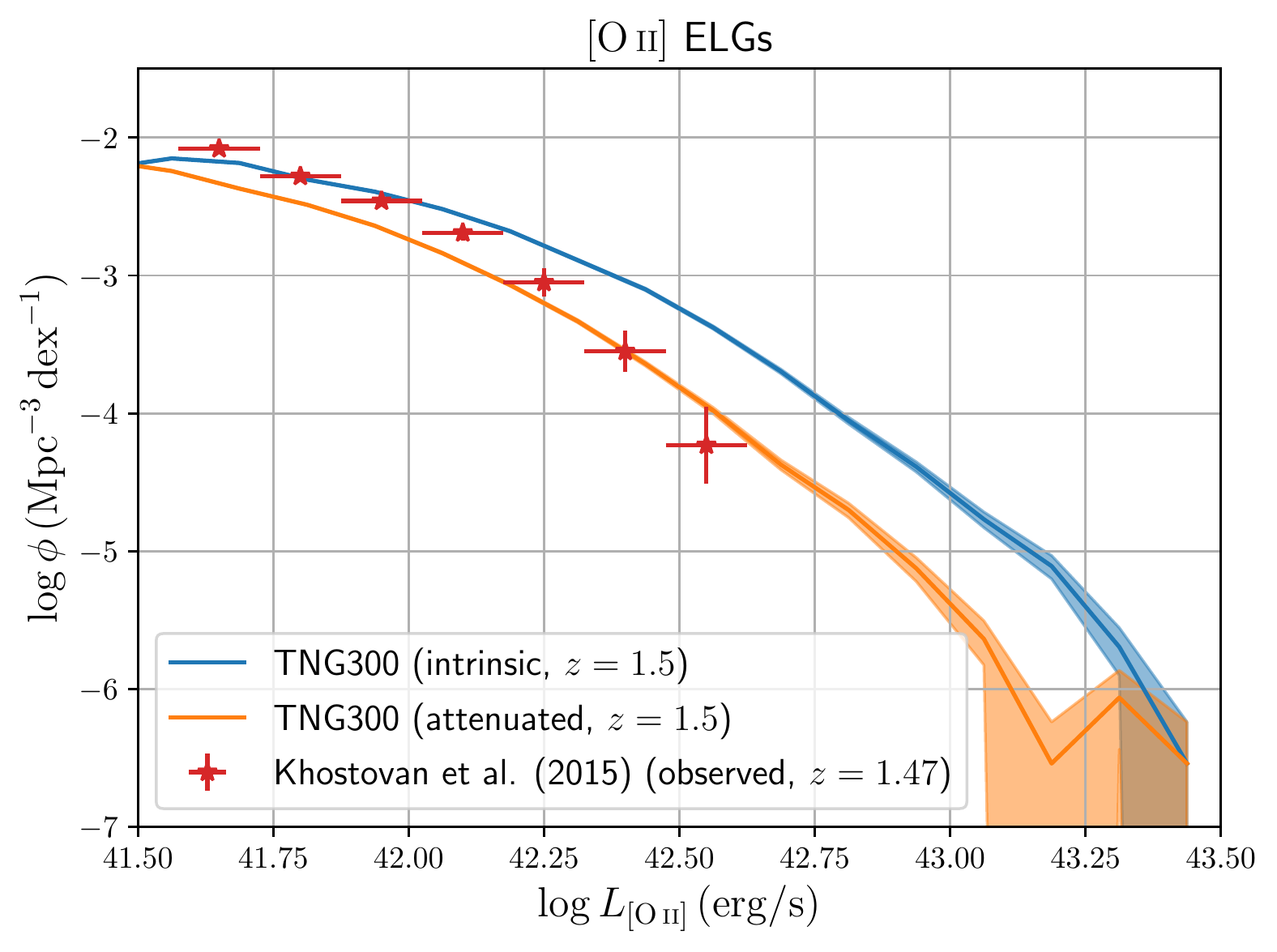}
  \caption{Luminosity functions of \HA ELGs (upper panel)
  and \OII ELGs (lower panel).
  Orange solid lines show the results with the dust attenuation effect considered but
  blue solid lines correspond to the results of intrinsic luminosities, i.e.,
  without dust attenuation.
  The error bars are determined as Poisson error.
  We also show the results of observations of \citet{Sobral2013} for \HA ELGs
  and \citet{Khostovan2015} for \OII ELGs.
  Note that the values presented in \citet{Sobral2013} are after dust correction.
  We convert them to observed luminosities with the dust extinction magnitude
  $A_{\mathrm{H}\alpha} = 1 \, \mathrm{mag}$.
  The values in \citet{Khostovan2015} are before dust correction, and thus we use the raw values.}
  \label{fig:LF}
\end{figure}

\subsection{Galactic properties of ELGs}
\label{sec:properties_ELGs}
In order to verify our mock ELG catalogues, we compare galactic properties
with observational estimates.

First, we measure the relation between star formation rate (SFR)
and emission line luminosities.
Since the nebular emission is driven by massive stars,
the line luminosity is an indicator of star formation activity and
the relation between SFR and line luminosities is well approximated
with a power-law function \citep{Kennicutt1998a}.
In hydrodynamical simulations, parameters related with star formation are calibrated
so that the Kennicutt--Schmidt law \citep{Kennicutt1998b} should be reproduced \citep{Schaye2008,Pillepich2018a}.
In this sense, measuring the relation can be regarded as a sanity check
of the mock ELG catalogues.
In Figure~\ref{fig:SFR_Lum}, we show the SFR and \HA and \OII line luminosity relations
measured from mock ELG catalogues
along with observational estimates from \citet{Davies2016} assuming the initial
mass function of \citet{Chabrier2003}:
\begin{align}
  \label{eq:SFR_HA}
  \mathrm{SFR}_{\text{\HA}} (\Msun \, \mathrm{yr}^{-1})
  &= \frac{L_{\text{\HA}}}{8.30 \times 10^{40} \, \mathrm{erg} \, \mathrm{s}^{-1}} , \\
  \label{eq:SFR_OII}
  \mathrm{SFR}_{\text{\OII}} (\Msun \, \mathrm{yr}^{-1})
  &= \frac{L_{\text{\OII}}}{9.67 \times 10^{40} \, \mathrm{erg} \, \mathrm{s}^{-1}} ,
\end{align}
where $L_{\text{\HA}}$ and $L_{\text{\OII}}$ are \HA and \OII line luminosities
with dust attenuation effect considered, respectively.
Though there are scatters at the faint end,
line luminosities and SFR are in power-law relations
and consistent with the result of \citet{Davies2016}.

Next, we consider luminosity functions of ELGs.
In Figure~\ref{fig:LF}, we compare the luminosity functions of \HA and \OII ELGs
with the observations of \citet{Sobral2013} for \HA ELGs
and \citet{Khostovan2015} for \OII ELGs.
Since the line luminosity is subject to dust extinction,
to obtain consistent results with observations,
we need to take into account the dust attenuation effect,
which was described in Section~\ref{sec:SPS}.
Otherwise, the luminosity function would be overestimated.
After correcting the dust attenuation effect, the measured luminosity function
is consistent with the observations within $0.5 \, \mathrm{dex}$.
Thus, we conclude our mock ELG catalogues are reliable at the reasonable level.
The effect of resolution of the simulations for luminosity functions
is addressed in Appendix~\ref{sec:convergence}.

\begin{figure}
  \includegraphics[width=\columnwidth]{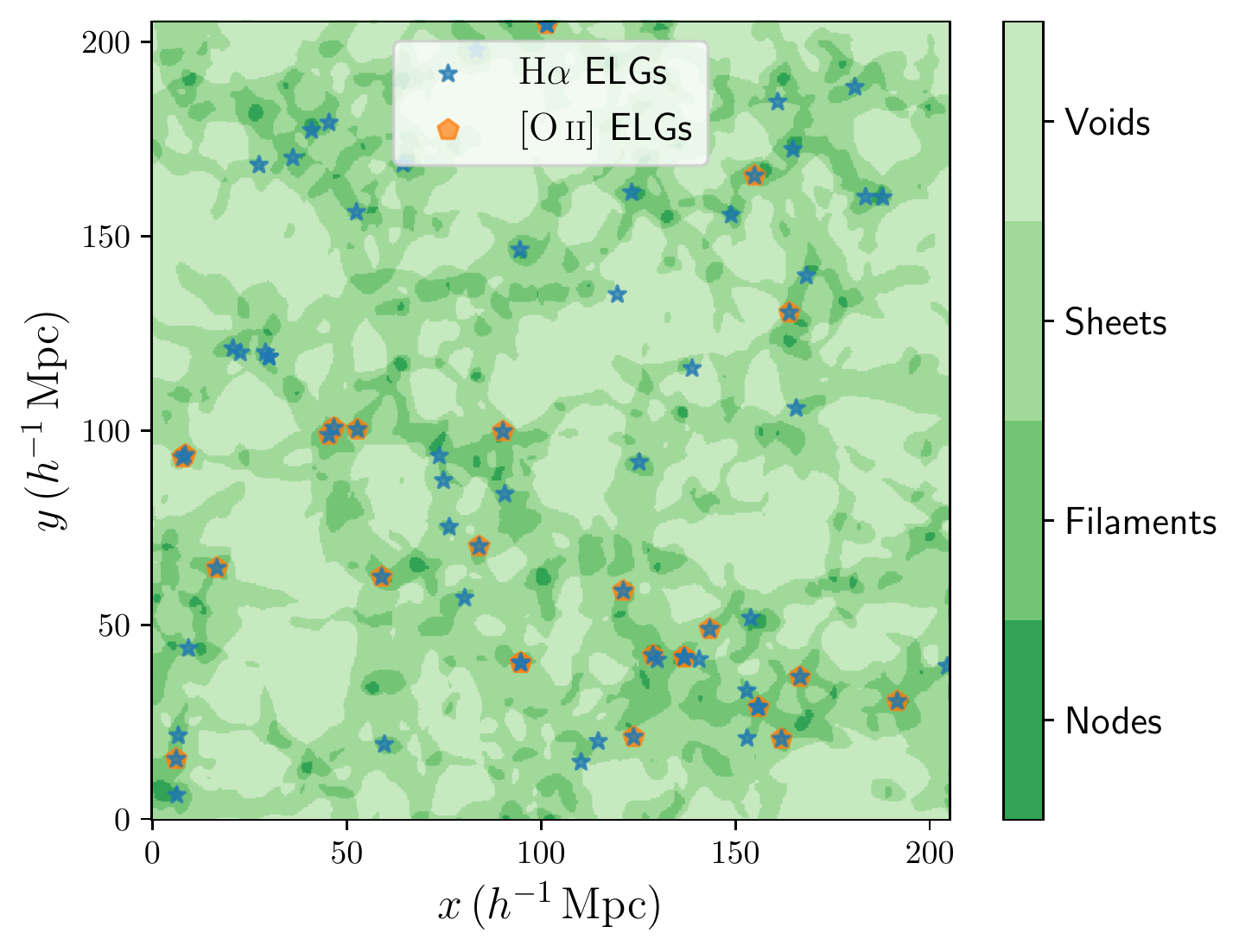}
  \caption{Classification of cosmic web structures for a slice of $0.40 \, \hMpc$.
  The entire simulation volume is divided into four classes:
  nodes, filaments, sheets, and voids.
  The positions of \HA and \OII ELGs are also shown as blue and orange symbols, respectively.}
  \label{fig:web}
\end{figure}

\begin{figure}
  \includegraphics[width=\columnwidth]{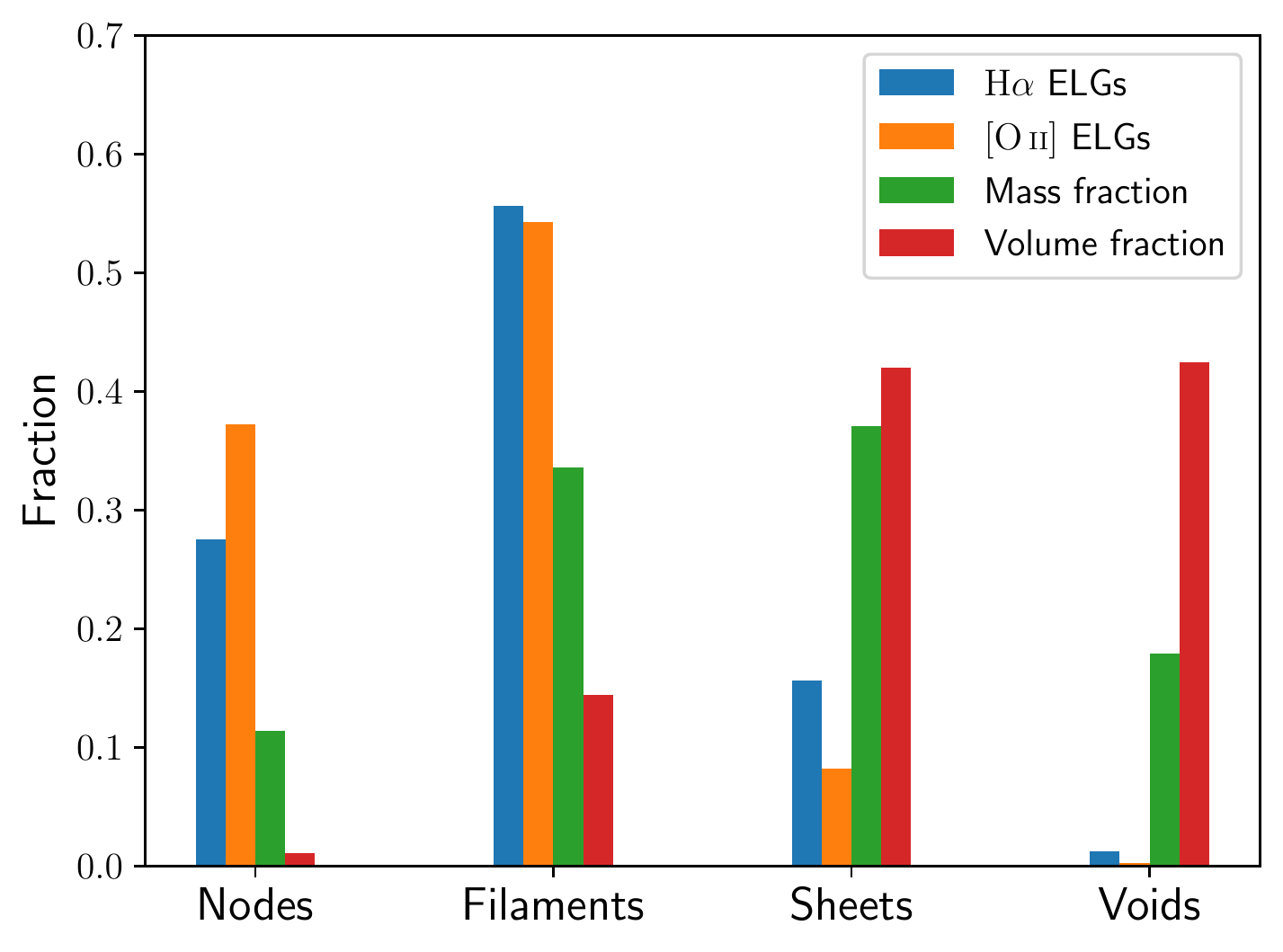}
  \caption{Fractions of \HA and \OII ELGs
  with respect to the environments: nodes, filaments, sheets, and voids.
  For comparison, the mass and volume fractions are also shown.}
  \label{fig:fractions_web}
\end{figure}

\begin{figure}
  \includegraphics[width=\columnwidth]{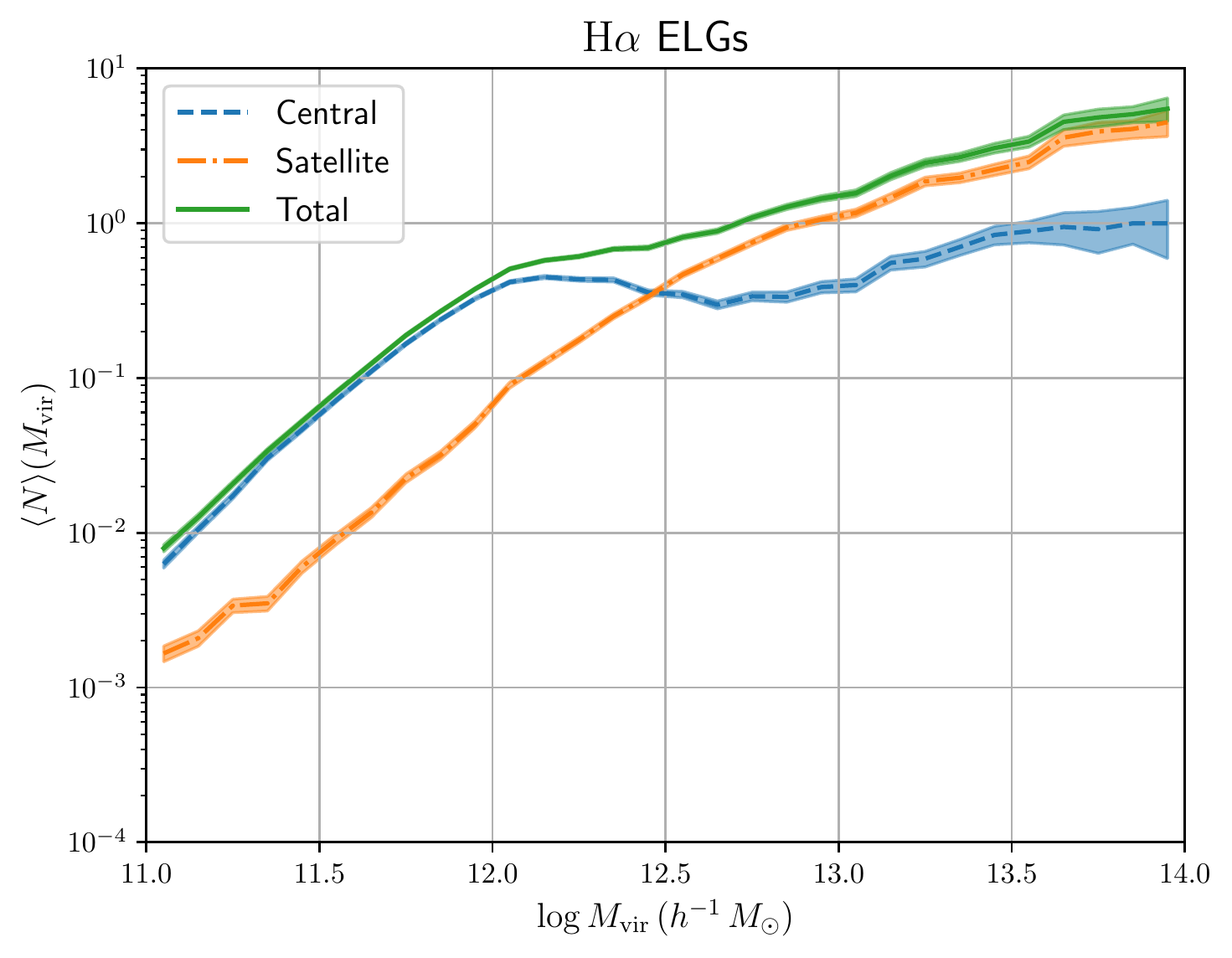}
  \includegraphics[width=\columnwidth]{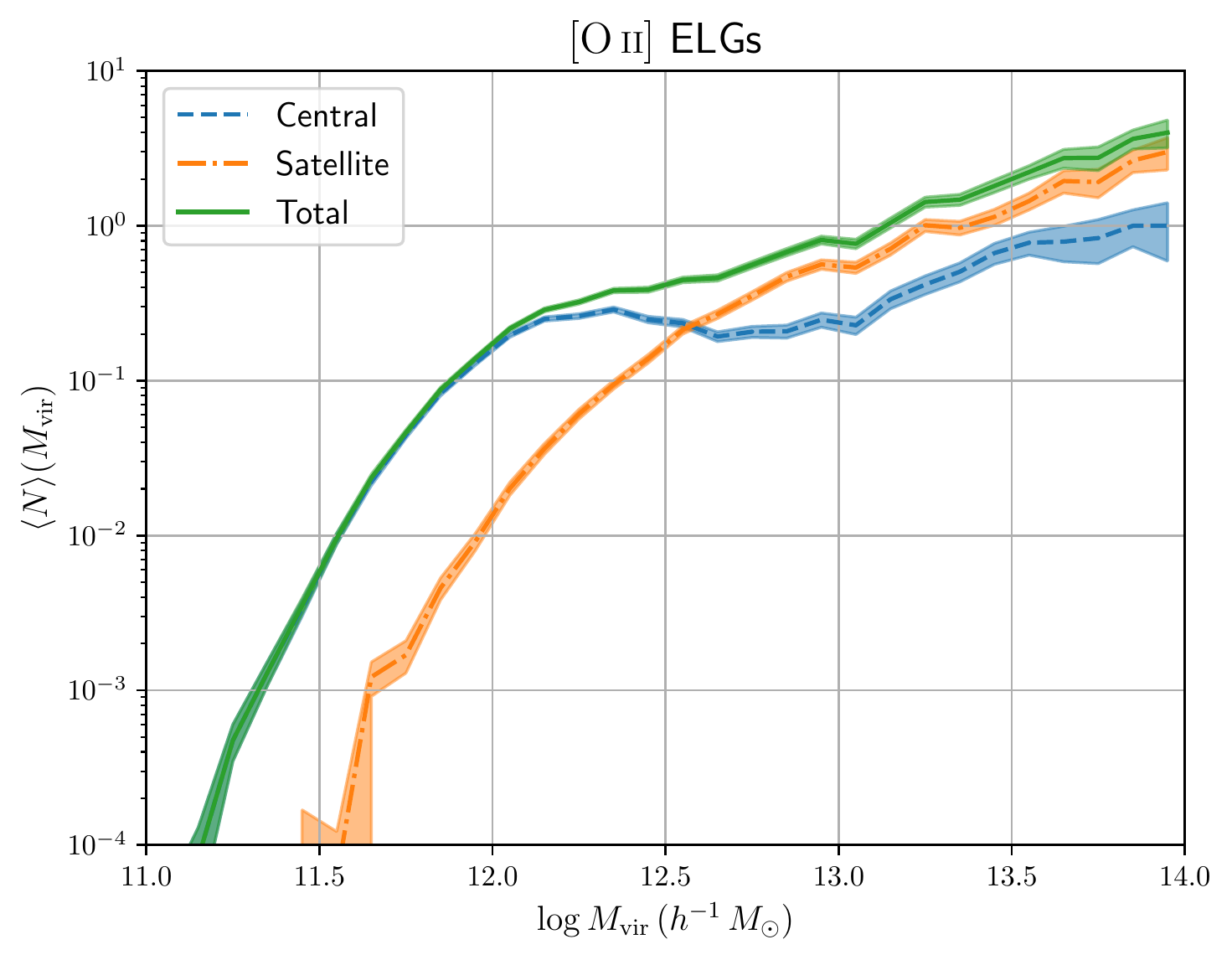}
  \caption{HODs of \HA ELGs (upper panel) and \OII ELGs (lower panel)
  measured from simulations.
  Each ELG is classified as a central galaxy or a satellite galaxy
  based on whether the subhalo hosting the ELG is the subhalo or the host halo,
  which is derived with the \texttt{SubFind} algotirhm \citep{Springel2001}.
  HODs of central and satellite ELGs are also shown as
  separate dashed and dot-dashed lines, respectively.
  To construct ELG samples, the luminosity threshold
  ($L > 10^{42} \, \mathrm{erg} \, \mathrm{s}^{-1}$) is applied,
  which corresponds to the number densiteis of 
  $n_{\text{\HA}} = 3.627 \times 10^{-3} \, (h^{-1} \, \mathrm{Mpc})^{-3}$
  and $n_{\text{\OII}} = 1.277 \times 10^{-3} \, (h^{-1} \, \mathrm{Mpc})^{-3}$
  for \HA and \OII ELGs, respectively.
  The shaded regions correspond to Poisson errors.}
  \label{fig:HOD}
\end{figure}

\begin{figure}
  \includegraphics[width=\columnwidth]{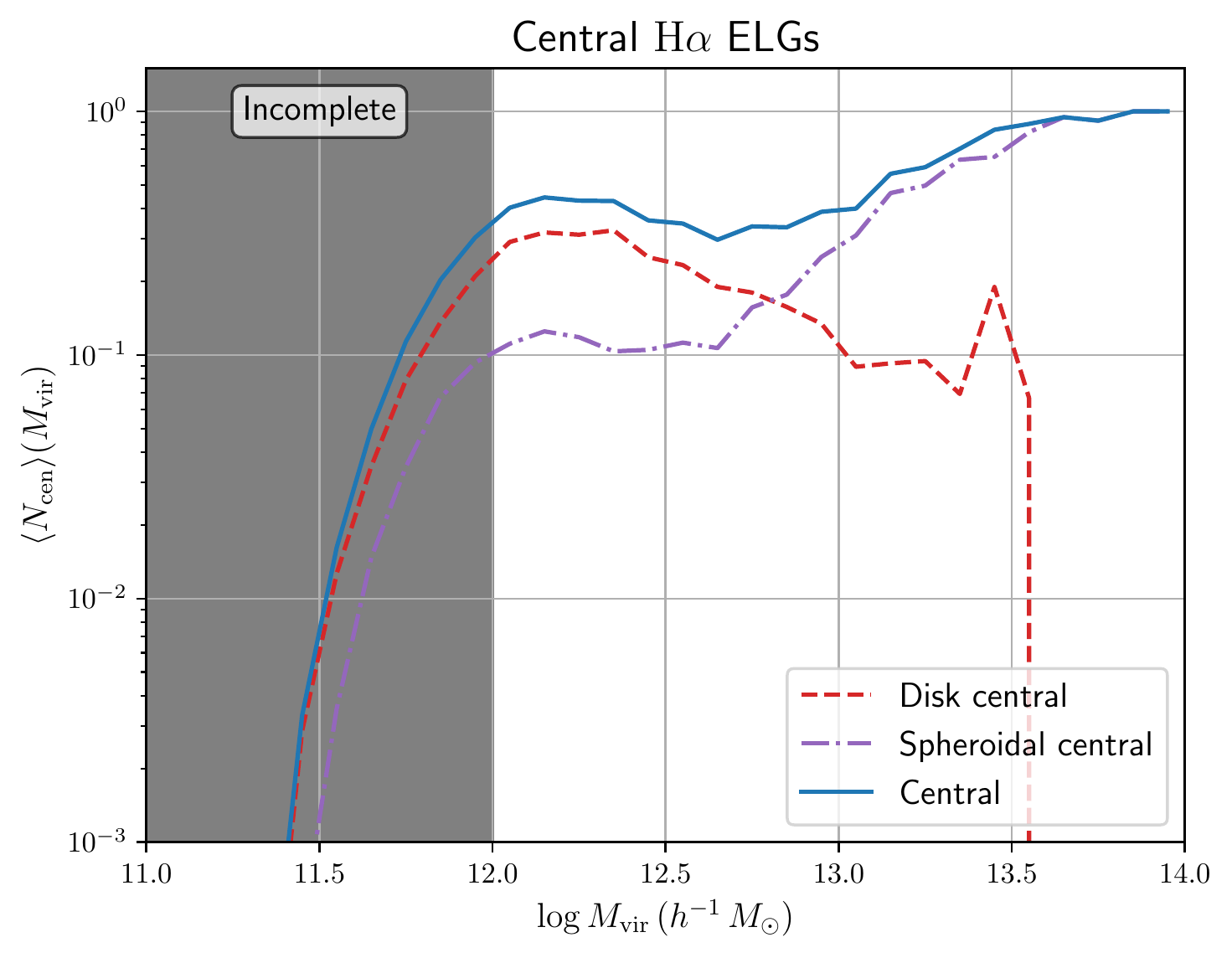}
  \includegraphics[width=\columnwidth]{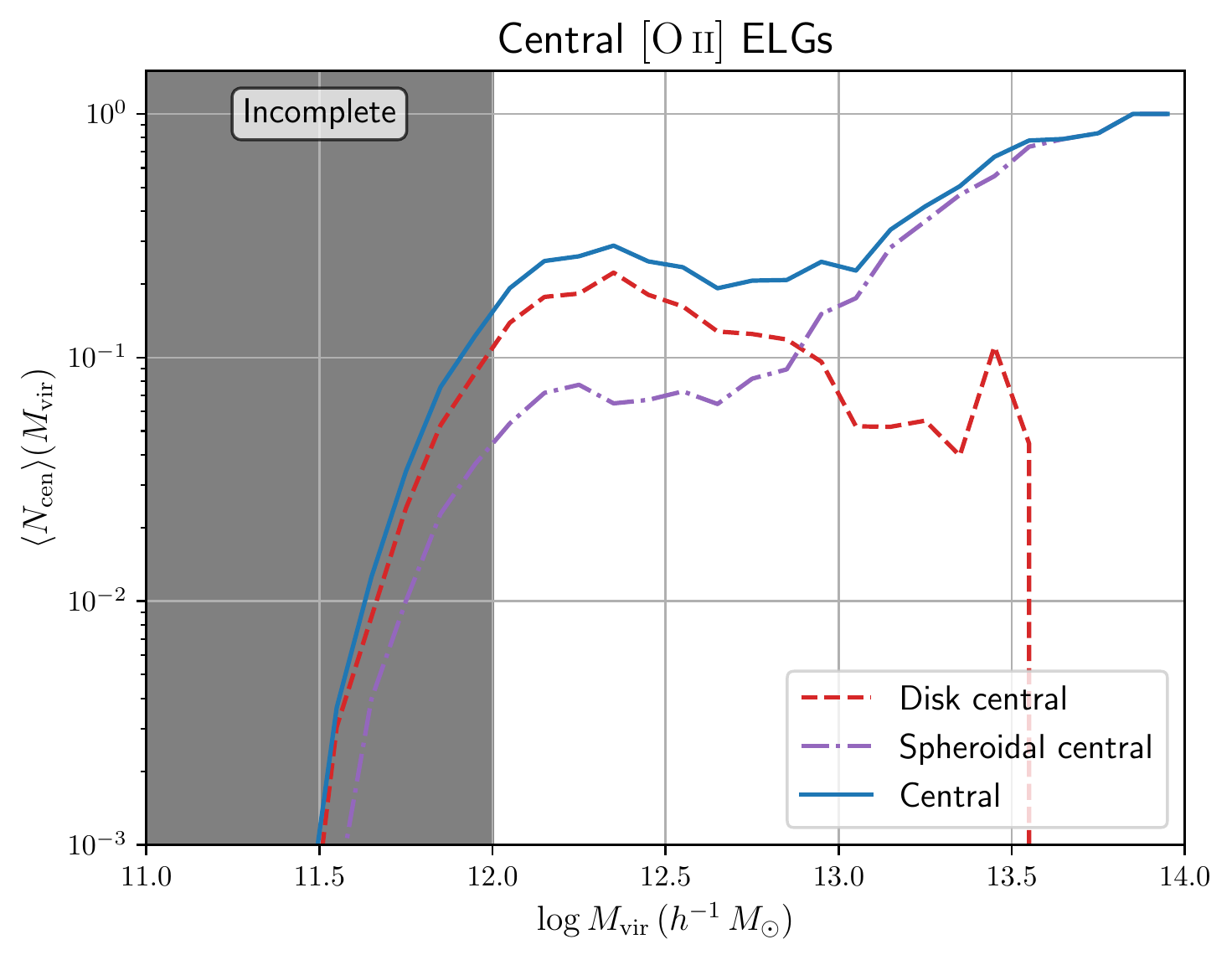}
  \caption{HODs of central \HA ELGs (upper panel) and \OII ELGs (lower panel).
  The central ELGs are split into disk centrals and spheroidal centrals
  according to the bulge over total mass ratio.
  Since some of halos with masses $< 10^{12} \, \hMsun$, which are displayed as grey regions
  in the figures, do not contain enough number of stellar particles to determine the bulge mass,
  and thus the total number of centrals at this mass range is incomplete.
  }
  \label{fig:HOD_disk_sph}
\end{figure}

\subsection{Cosmic web environments around ELGs}
Since ELGs are star-forming young populations,
we expect that the majority of ELGs is infalling
towards massive halos along filamentary structures \citep{GonzalezPerez2020}.
We investigate where mock ELGs are found in terms of cosmic web structures
\citep{Hahn2007a,Hahn2007b,Cautun2014,Libeskind2018}.
First, we divide the entire simulation volume into four classes:
nodes, filaments, sheets, and voids.
For this purpose, we define the tidal tensor of the gravitational potential:
\begin{equation}
  T_{ij} (\bm{x}) \equiv \pdv{}{x_i}{x_j} \tilde{\phi} (\bm{x}),
\end{equation}
where $\bm{x}$ is the comoving coordinates,
$\tilde{\phi} (\bm{x}) = \phi (\bm{x}) / (4 \pi G \bar{\rho}_\mathrm{m})$
is the normalized gravitational potential,
$\phi (\bm{x})$ is the gravitational potential,
$G$ is the gravitational constant,
and $\bar{\rho}_\mathrm{m}$ is the mean matter density at the present Universe.
The potential can be computed via the Poisson equation:
\begin{equation}
  \nabla^2 \tilde{\phi} (\bm{x}) = \delta (\bm{x}) ,
\end{equation}
where $\delta (\bm{x})$ is the density contrast.
The density contrast is computed with regular grids
by assigning matter particles with cloud-in-cell (CIC) algorithm.
The number of grid is $512$ on a side and the corresponding grid size is
$205 \, \hMpc /512 = 0.40 \, \hMpc$.
In order to avoid small-scale transients, we apply Gaussian smoothing
with the filter function:
\begin{equation}
  W_\mathrm{G} (x) = \frac{1}{(2 \pi \sigma_\mathrm{G})^{3/2}}
  \exp \left[ - \frac{x^2}{2\sigma_\mathrm{G}^2}\right],
\end{equation}
where the filtering scale is $\sigma_\mathrm{G} = 2 \, \hMpc$.
Then, we deconvolve the aliasing effect due to CIC assignment \citep{Jing2005}.
Next, we compute the eigenvalues of the tidal tensor for each position of regular grids,
which are denoted as $(\lambda_1, \lambda_2, \lambda_3)$ in an ascending order,
and the position is classified according
to the number of eigenvalues which exceed the threshold value $\lambda_\mathrm{th}$:
\begin{itemize}
  \item Nodes: $\lambda_\mathrm{th} < \lambda_1 < \lambda_2 < \lambda_3$,
  \item Filaments: $\lambda_1 < \lambda_\mathrm{th} < \lambda_2 < \lambda_3$,
  \item Sheets: $\lambda_1 < \lambda_2 < \lambda_\mathrm{th} < \lambda_3$,
  \item Voids: $\lambda_1 < \lambda_2 < \lambda_3 < \lambda_\mathrm{th}$.
\end{itemize}
We adopt the threshold value $\lambda_\mathrm{th} = 0.2$,
which reproduces the visual cosmic web structures \citep{Forero-Romero2009}.
In Figure~\ref{fig:web}, we illustrate the cosmic web classification of the slice
with the thickness of $0.40 \, \hMpc$
together with the positions of \HA and \OII ELGs.
Here, ELGs are selected
with the luminosity cut $L > 10^{42} \, \mathrm{erg} \, \mathrm{s}^{-1}$.
Visually, the distribution of ELGs traces the LSS,
and most of ELGs are located at nodes or filaments and
ELGs at voids are quite rare,
which is consistent with the results of semi-analytic simulations
\citep{GonzalezPerez2020} and hydrodynamical simulations \citep{Hadzhiyska2021}.
In Figure~\ref{fig:fractions_web}, we show the fractions of ELGs
in terms of the four categories.
The fractions for ELGs clearly show tendency different from mass and volume fractions.
More than $50\%$ of ELGs are found in filaments,
the rest is in nodes or sheets, and almost no ELGs are found in voids.
The results indicate that a large fraction of ELGs is the population of
infalling galaxies towards the gravitational potential well.

\subsection{Halo occupation distribution of mock ELGs}
\label{sec:HOD_measurement}
Next, we measure the HODs of \HA and \OII ELGs directly from simulations.
Since the halo catalogues of IllustrisTNG simulations constructed
with the \texttt{SubFind} algorithm \citep{Springel2001} are provided,
the information of mass and relation between host halos and subhalos is readily accessible.
The most massive subhalo, which is located at the centre of the halo in most of cases,
is labeled as a central galaxy and all the other subhalos are satellite galaxies.
Figure~\ref{fig:HOD} shows the measured HODs with the line luminosity cut
$L > 10^{42} \, \mathrm{erg} \, \mathrm{s}^{-1}$.
In contrast to LRGs,
the HOD of centrals cannot be well described by the step-like function
but there are two components:
a log-normal function which peaks around the mass of $10^{12} \, \hMsun$
and a step-like function which gradually reaches unity at the massive end.
This overall feature is consistent with results of semi-analytic simulations \citep{GonzalezPerez2018}
and hydrodynamical simulations \citep{Hadzhiyska2021},
and both of the results also exhibit the peak of the central HOD at $10^{12} \, \hMsun$.
Though the non-vanishing central HOD at the massive end
is not incorporated in some HOD models for ELGs
\citep[see, e.g.,][]{Avila2020},
the similar feature can be seen in semi-analytic simulations \citep{GonzalezPerez2018}.
One of the possible reasons for the non-vanishing central HODs
is that any magnitude cuts or colour selections
\citep[see, e.g.,][]{Raichoor2017} are not applied
to construct our mock ELG catalogues and such selection may exclude
central ELGs hosted by massive halos.
The target selection of ELG candidates with our mock galaxy suite
will be addressed in a future work.

\begin{table}
  \caption{Range of HOD parameters adopted for the Zheng and Geach models.}
  \label{tab:parameter_range_HOD}
  \begin{tabular}{cc}
    \hline
    Parameter & Range \\
    \hline \hline
    \multicolumn{2}{c}{Zheng HOD} \\
    \hline
    $\sigma_{\log M}$, $\alpha$ & $[0, \infty)$ \\
    $\log M_\mathrm{min}$, $\log M_0$, $\log M_1$
    & $[8, 16] \, \log \hMsun$ \\
    \hline
    \multicolumn{2}{c}{Geach HOD} \\
    \hline
    $F_\mathrm{c}$ & $[0, 1]$ \\
    $F_\mathrm{s}$, $\sigma_{\log M_\mathrm{c1}}$, $\sigma_{\log M_\mathrm{c2}}$,
    $\sigma_{\log M_\mathrm{s}}$, $\alpha$ & $[0, \infty)$ \\
    $\log M_\mathrm{c1}$, $\log M_\mathrm{c2}$, $\log M_\mathrm{s}$
    & $[8, 16] \, \log \hMsun$ \\
    \hline
  \end{tabular}
\end{table}

\begin{figure*}
	\includegraphics[width=\textwidth]{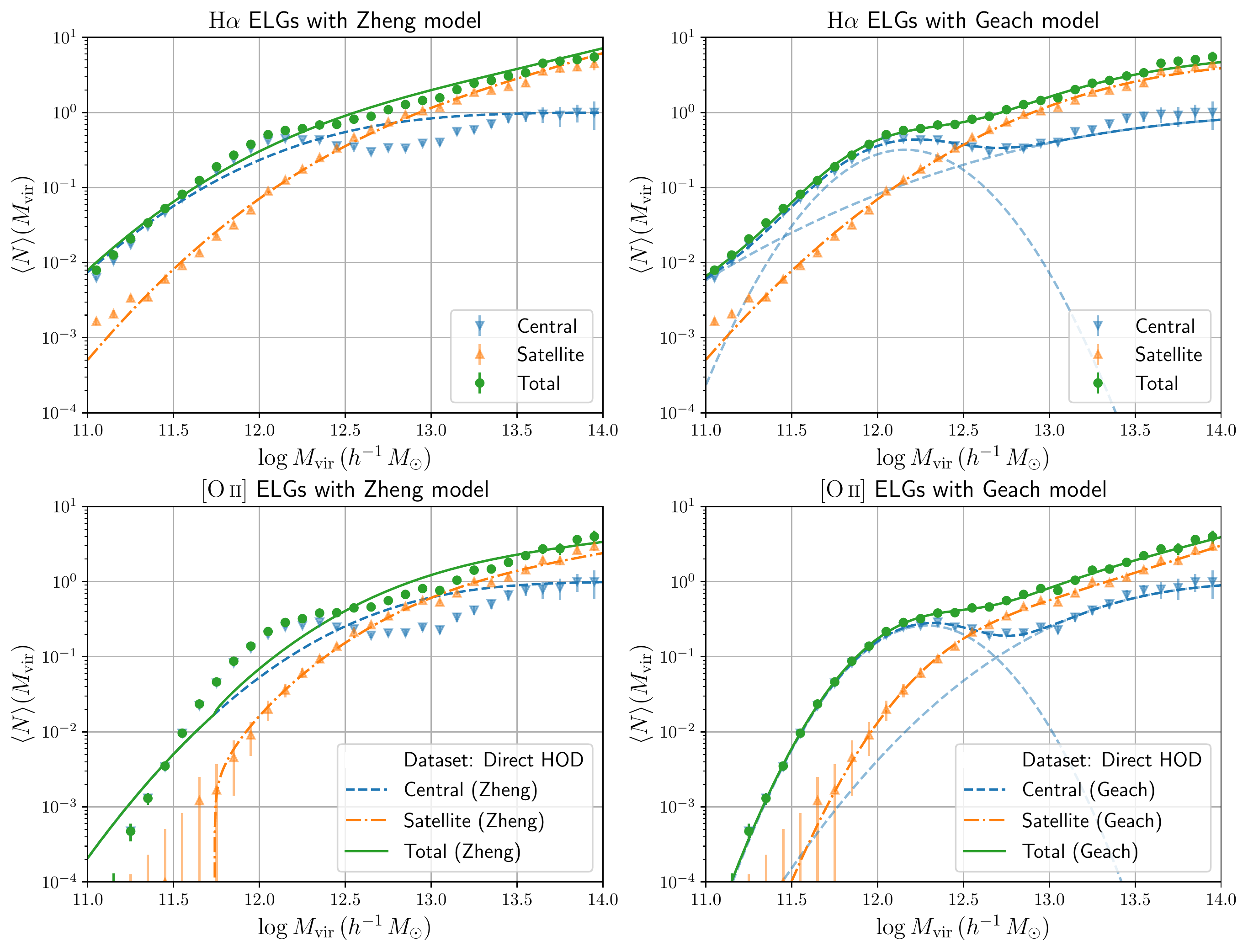}
  \caption{Best-fit HODs with the Zheng (left panels) and Geach (right panels)
  models for \HA ELGs (upper panels) and \OII ELGs (lower panel).
  For the Geach model, the log-normal and step-like terms for central HODs are
  displayed as separate dashed lines.}
  \label{fig:bestfit_HOD}
\end{figure*}

In order to investigate the origin of the peak of the central HODs,
in \citet{GonzalezPerez2018}, the central ELGs are split into two populations, disk centrals
and spheroidal centrals, based on a bulge over total mass ratio,
and the former and latter corresponds to the log-normal and step-like function feature,
respectively.
The disk centrals undergo active star formation and
are infalling towards the massive halos along filaments
and considered as a majority of ELGs.
On the other hand, spheroidal centrals are located close to the centre of the hosting halos
and thus are expected to be quenched and not to be observed as ELGs.
However, in the dense region such as the centre of halos,
gas as fuel of star formation is replenished and star formation is kept active.
Therefore, such spheroidal centrals are bright in the line emission.
To verify this picture, we perform the analysis similar
to that of \citet{GonzalezPerez2018} for our mock \HA and \OII ELGs.
Each stellar particle within a galaxy is classified as disk or bulge component
according to the circularity parameter $\epsilon$ \citep{Abadi2003}.
We make use of the supplementary catalogue of stellar circularity of \citet{Genel2015}
and define the bulge mass fraction as the double of the fractional mass of stars with $\epsilon < 0$.
Then, if the bulge mass fraction is larger than $0.5$, the galaxy is classified as a ``spheroidal galaxy''
and otherwise as a ``disk galaxy''.
Figure~\ref{fig:HOD_disk_sph} shows the central HODs of \HA and \OII ELGs with central ELGs
split into disk or spheroidal centrals.
Though the decomposition is not possible
for some of low mass halos ($M_\mathrm{vir} < 10^{12} \, \hMsun$)
due to the lack of the resolution,
the transition from disk centrals to spheroidal centrals towards the massive end
is clearly seen and the peak around $10^{12} \, \hMsun$ is dominated by disk centrals.
This overall trend is consistent with the results of \citet{GonzalezPerez2018}.
For satellite HODs, previous studies \citep{Geach2012,Contreras2013,Avila2020}
suggest that the slope should be around $0.8\text{--}1$ but
for our mock ELGs, a smaller slope ($\lesssim 0.7$) is preferred.
That might indicate rapid quenching once ELGs infall to the halos in hydrodynamical simulations.
Though these features are specific to our mock ELG catalogues,
they appear at the massive end.
Hence, the number of such massive halos is so small
that the deviation does not have impacts on clustering statistics.

\begin{figure}
  \includegraphics[width=\columnwidth]{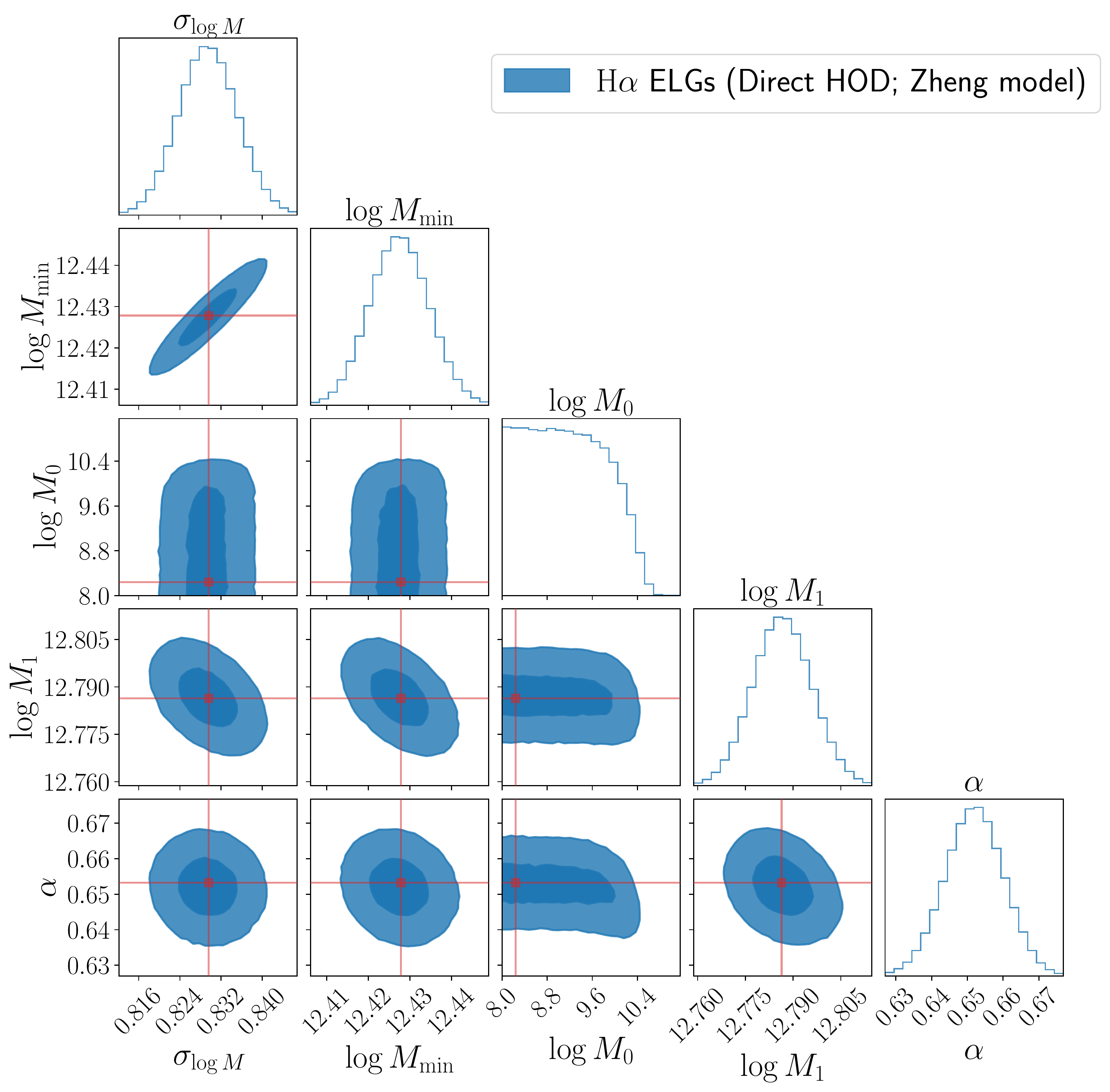}
  \caption{Constraints of HOD parameters for \HA ELGs with the Zheng model
  by fitting the direct HOD.
  The thick (thin) blue regions correspond to $1\sigma$ ($2\sigma$) confidence level.
  The red lines indicate best-fit values. The unit of mass parameters
  ($M_\mathrm{min}$, $M_0$, $M_1$) is $\hMsun$.}
  \label{fig:Zheng_HA_HOD_triangle}
\end{figure}

\begin{figure}
  \includegraphics[width=\columnwidth]{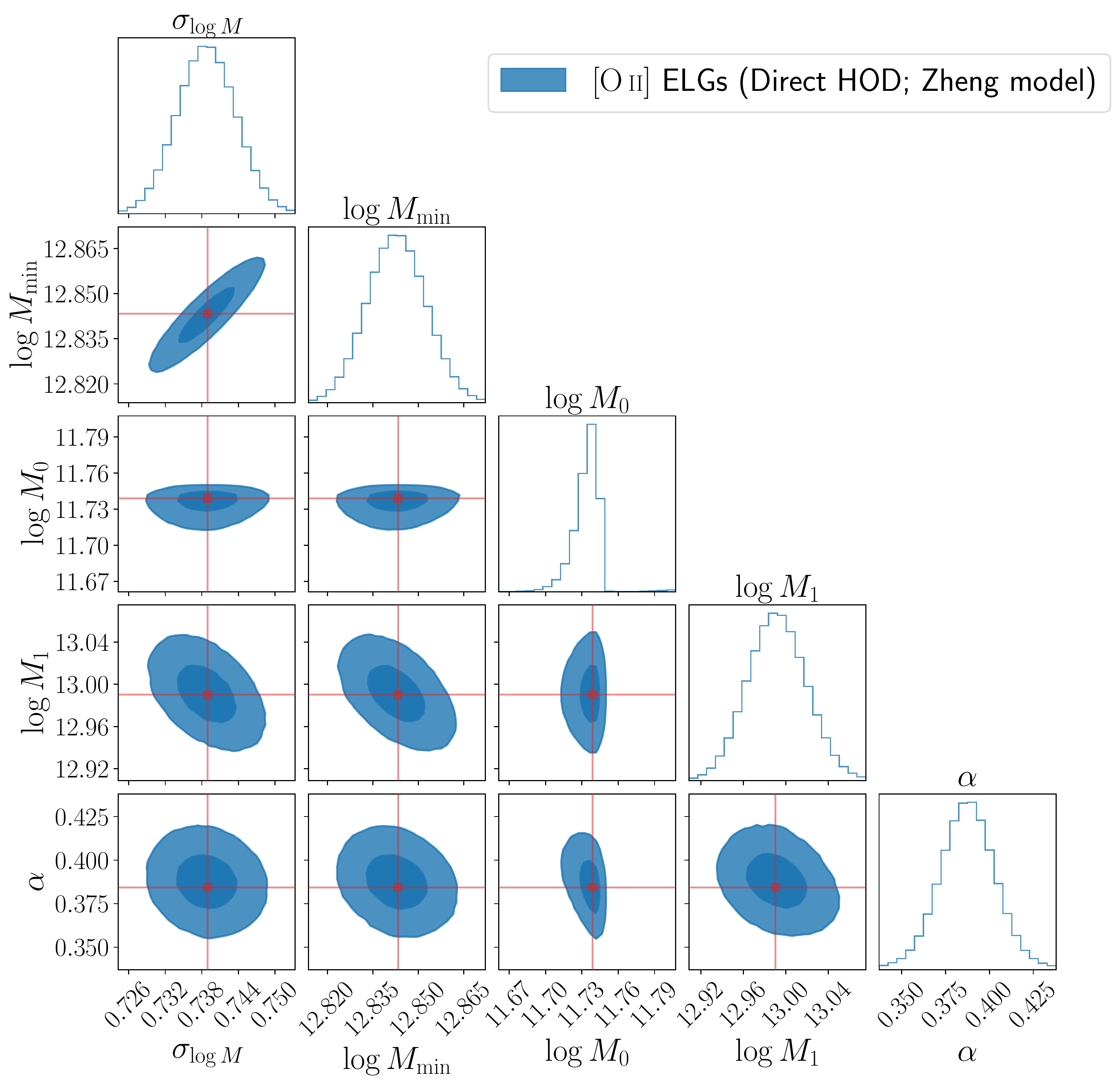}
  \caption{Same as Figure~\ref{fig:Zheng_HA_HOD_triangle}
  but for \OII ELGs with the Zheng model.}
  \label{fig:Zheng_OII_HOD_triangle}
\end{figure}

\begin{figure*}
  \includegraphics[width=\textwidth]{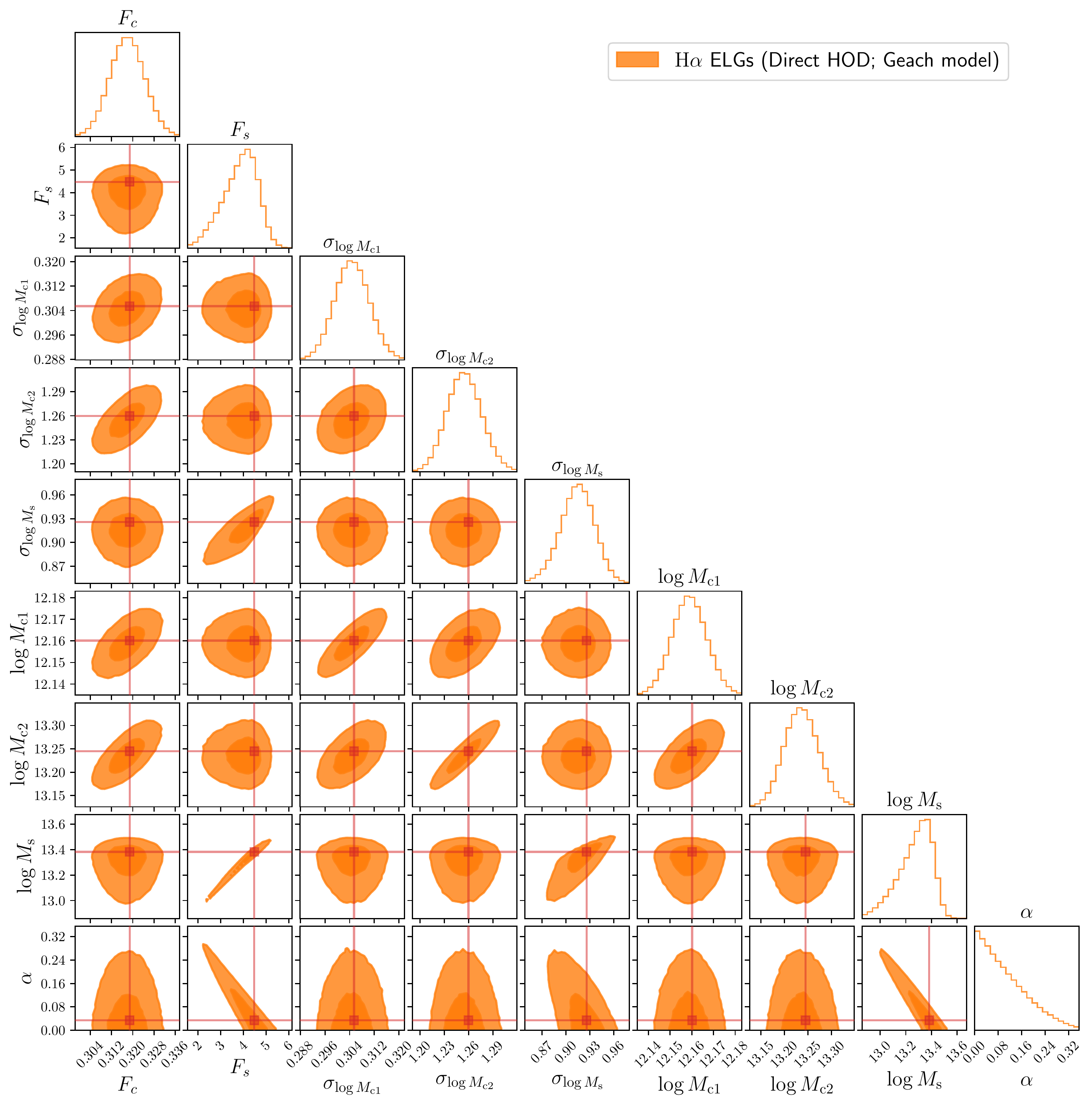}
  \caption{Same as Figure~\ref{fig:Zheng_HA_HOD_triangle}
  but for \HA ELGs with the Geach model.
  The unit of mass parameters
  ($M_\mathrm{c1}$, $M_\mathrm{c2}$, $M_\mathrm{s}$) is $\hMsun$.}
  \label{fig:Geach_HA_HOD_triangle}
\end{figure*}

\begin{figure*}
  \includegraphics[width=\textwidth]{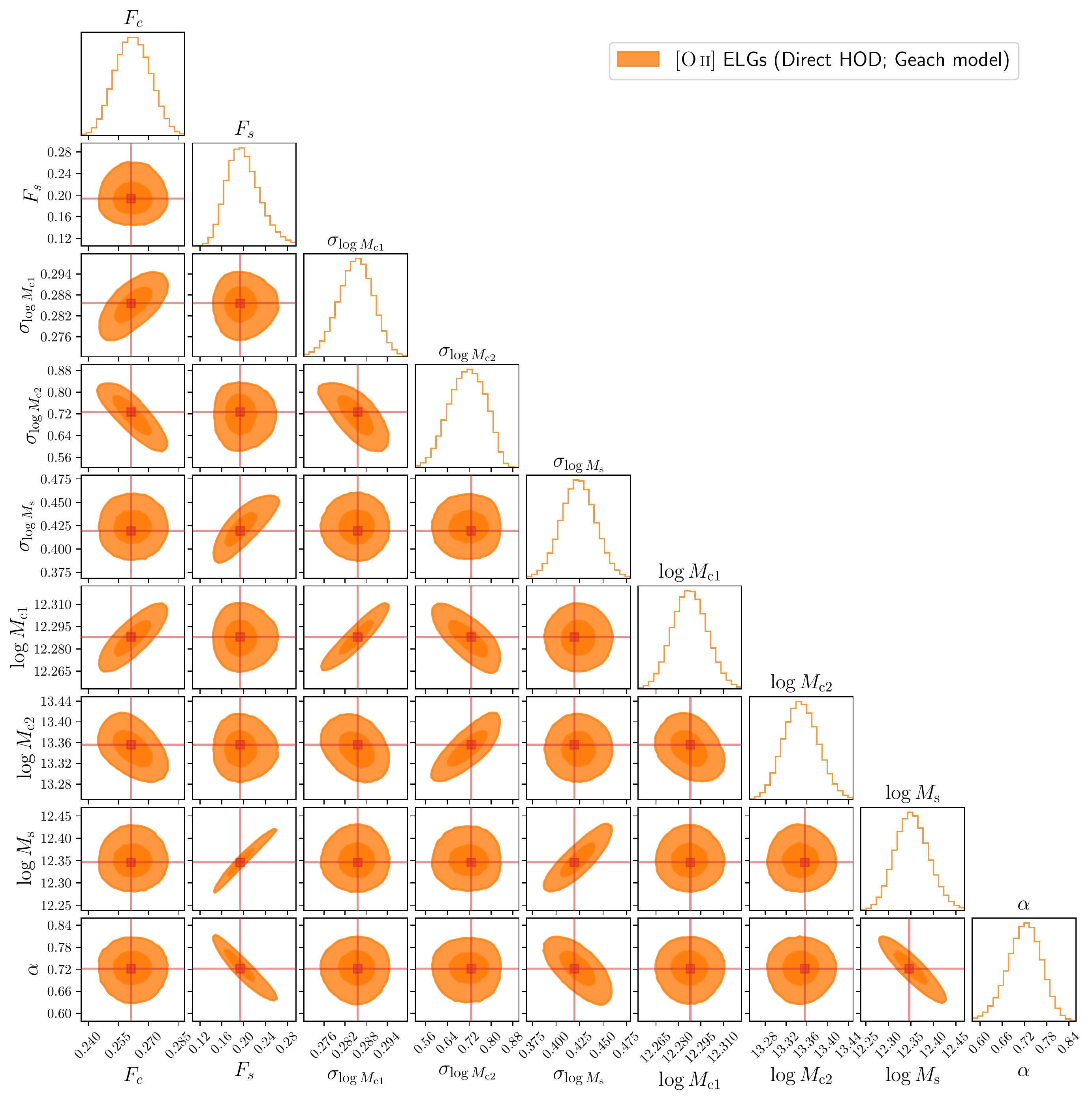}
  \caption{Same as Figure~\ref{fig:Zheng_HA_HOD_triangle}
  but for \OII ELGs with the Geach model.
  The unit of mass parameters
  ($M_\mathrm{c1}$, $M_\mathrm{c2}$, $M_\mathrm{s}$) is $\hMsun$.}
  \label{fig:Geach_OII_HOD_triangle}
\end{figure*}

\section{Inference of HOD parameters}
\label{sec:inference_HOD}

\subsection{Inference with direct HOD}
\label{sec:direct_HOD}
In this section, we directly fit HOD parameters of the Zheng and Geach models with measured HODs
and investigate which model fits the measured HODs well.
In order to avoid the confusion of terminology, hereafter we refer to the HOD
measured from simulations as ``direct HOD'' data set.
Hereafter, the mock ELG sample consists of ELGs
with the luminosity cut $L > 10^{42} \, \mathrm{erg} \, \mathrm{s}^{-1}$,
which is similar to the flux limit in the PFS survey.
The resultant \HA and \OII ELG number densities are
$n_{\text{\HA}} = 3.627 \times 10^{-3} \, (h^{-1} \, \mathrm{Mpc})^{-3}$
and $n_{\text{\OII}} = 1.277 \times 10^{-3} \, (h^{-1} \, \mathrm{Mpc})^{-3}$, respectively.
The likelihood of the direct HOD plus the ELG number density,
$\mathcal{L}_{\mathrm{HOD} + n_\mathrm{g}}$, is given as
\begin{align}
  \log \mathcal{L}_{\mathrm{HOD} + n_\mathrm{g}} &=
  -\frac{1}{2} \left( \chi^2_\mathrm{cen} + \chi^2_\mathrm{sat} + \chi^2_{n_\mathrm{g}} \right)
  +\text{const.} , \\
  \label{eq:chi2_cen}
  \chi^2_\mathrm{cen} &=
  \sum_i \frac{\left( \langle \hat{N}_\mathrm{cen} \rangle (M_i) -
  \langle N_\mathrm{cen} \rangle (M_i) \right)^2}{\sigma_\mathrm{cen}^2 (M_i)}, \\
  \label{eq:chi2_sat}
  \chi^2_\mathrm{sat} &=
  \sum_i \frac{\left( \langle \hat{N}_\mathrm{sat} \rangle (M_i) -
  \langle N_\mathrm{sat} \rangle (M_i) \right)^2}{\sigma_\mathrm{sat}^2 (M_i)}, \\
  \chi^2_{n_\mathrm{g}} &= \frac{\left( \log_{10} \hat{n}_\mathrm{g} -
  \log_{10} n_\mathrm{g} \right)^2}
  {\sigma_{n_\mathrm{g}}^2} , \label{eq:chi2_ng}
\end{align}
where $\langle \hat{N}_\mathrm{cen,sat} \rangle$ is the measured mean occupation numbers of
central or satellite ELGs,
$\langle N_\mathrm{cen,sat} \rangle$ is the prediction based on the Geach or Zheng model, and
we also include the chi-square of the number density $\chi^2_{n_\mathrm{g}}$;
$\hat{n}_\mathrm{g}$ is the ELG number density measured in simulations and
$n_\mathrm{g}$ is the expected number density (Eq.~\ref{eq:HOD_ng}).
We estimate the errors of the occupation numbers, $\sigma_{n_\mathrm{cen}}$ and $\sigma_{n_\mathrm{cen}}$, assuming the Poisson distribution.
We set the error of the number density as $\sigma_{n_\mathrm{g}} = \tau |\log_{10} \hat{n}_\mathrm{g}|$
with $\tau = 0.03$ \citep[e.g.,][]{Okumura2021}.
The mass bin, which is labelled by the subscript $i$
in Eqs.~\eqref{eq:chi2_cen} and \eqref{eq:chi2_sat},
is logarithmically sampled in the range of
$[10^{11}, 10^{14}] \, h^{-1} \, \Msun$ with respect to virial mass with $30$ bins.
We assume flat priors for all the HOD parameters and the considered parameter range is given
in Table~\ref{tab:parameter_range_HOD}.
Since the central HOD of the Geach model is not bounded to one,
we also apply an additional prior that ensures
the central HOD is less than one for any halo mass.\footnote{In the practical implementation,
at every Markov chain step,
we check the central HOD is less than one and if not, the prior
is set to be zero.}
In order to compute the posterior distribution, we employ
Affine invariant Markov chain Monte-Carlo (MCMC) sampler \texttt{emcee} \citep{Foreman-Mackey2013}.
Figure~\ref{fig:bestfit_HOD} shows the best-fitting HOD with the Zheng and Geach models
and HOD parameter constraints are illustrated in
Figures~\ref{fig:Zheng_HA_HOD_triangle}, \ref{fig:Zheng_OII_HOD_triangle},
\ref{fig:Geach_HA_HOD_triangle}, and \ref{fig:Geach_OII_HOD_triangle}.
In general, the Geach model outperforms the Zheng model in terms of fitting the measured HOD.
The Zheng model can well fit the satellite HOD but fails to capture the two component feature of
the central HOD. In contrast, the Geach model takes into account the characteristic nature of
the central HOD by introducing the log-normal term and thus can explain the overall shape of the central HOD.
It also should be noted that most of the HOD parameters are robustly determined
and the marginalized posterior distributions are well approximated with normal distributions.

\subsection{HOD challenge with projected correlation functions}
In the real observations, HOD is not a direct observable
but can be inferred from statistics of ELG clustering.
The projected correlation function is
widely used to constrain HOD models.
We follow the procedure of the HOD parameter inference as in the practical measurements;
we constrain model HOD parameters with the Bayesian inference procedure
from the projected correlation function of mock ELGs.
In Section~\ref{sec:HOD_measurement}, we have measured the HOD directly from the simulations. We thus can compare the HOD constrained from clustering statistics with those directly measured, namely the direct HOD.
This analysis is a critical test for the HOD parameter inference
and we refer to it as an \textit{HOD challenge}.

The likelihood of the projected correlation function plus the ELG number density,
$\mathcal{L}_{w_\mathrm{p}+ n_\mathrm{g}}$,
is given as 
\begin{align}
  \log \mathcal{L}_{w_\mathrm{p} + n_\mathrm{g}} =&
  -\frac{1}{2} \chi^2_{w_\mathrm{p}} -\frac{1}{2} \log \det \mathrm{Cov} [w_\mathrm{p}]
  -\frac{1}{2} \chi^2_{n_\mathrm{g}} + \text{const} ,\\
  \chi^2_{w_\mathrm{p}} =& \sum_{i, j} (\hat{w}_\mathrm{p} (R_i) - \bar{w}_\mathrm{p} (R_i))
  \mathrm{Cov}^{-1}[w_\mathrm{p}] (R_i, R_j) \nonumber \\
  & \times (\hat{w}_\mathrm{p} (R_j) - \bar{w}_\mathrm{p} (R_j)) ,
\end{align}
where $\hat{w}_\mathrm{p}$ is the projected correlation function measured from the simulations,
$\bar{w}_\mathrm{p}$ is the model prediction of
the bin-averaged projected correlation functions (Eq.~\ref{eq:bin_averaged_wp}),
and $\chi^2_{n_\mathrm{g}}$ is given in equation (\ref{eq:chi2_ng}).
Note that the covariance matrix has parameter dependence and thus
the determinant of the covariance is no longer constant.
We assume the same parameter range as that used in Section~\ref{sec:direct_HOD}
(Table~\ref{tab:parameter_range_HOD})
and at every step of MCMC, we update the covariance matrix (Eq.~\ref{eq:wp_Cov}).
For the measurement of the projected correlation functions,
we employ \texttt{Corrfunc} \citep{Sinha2020}.
The transverse bin is logarithmically sampled in the range of $[0.05, 50] \, \hMpc$
with $15$ bins.
This scale range is matched with the one employed in the measurement
of the projected correlation function of \OII ELGs detected by Subaru HSC \citep{Okumura2021}.
We will address how the parameter inference is affected by including the scale cut in Appendix~\ref{sec:wp_cut}.
Note that we do not correct integral constraint \citep{Peebles1976} because
the random correlation is analytically computed under the periodic boundary condition
of the simulations and the effect is quite minor in the current setting.
Similarly to the inference from the direct HOD,
we employ \texttt{emcee} to compute the posterior distribution.

\begin{figure}
  \includegraphics[width=\columnwidth]{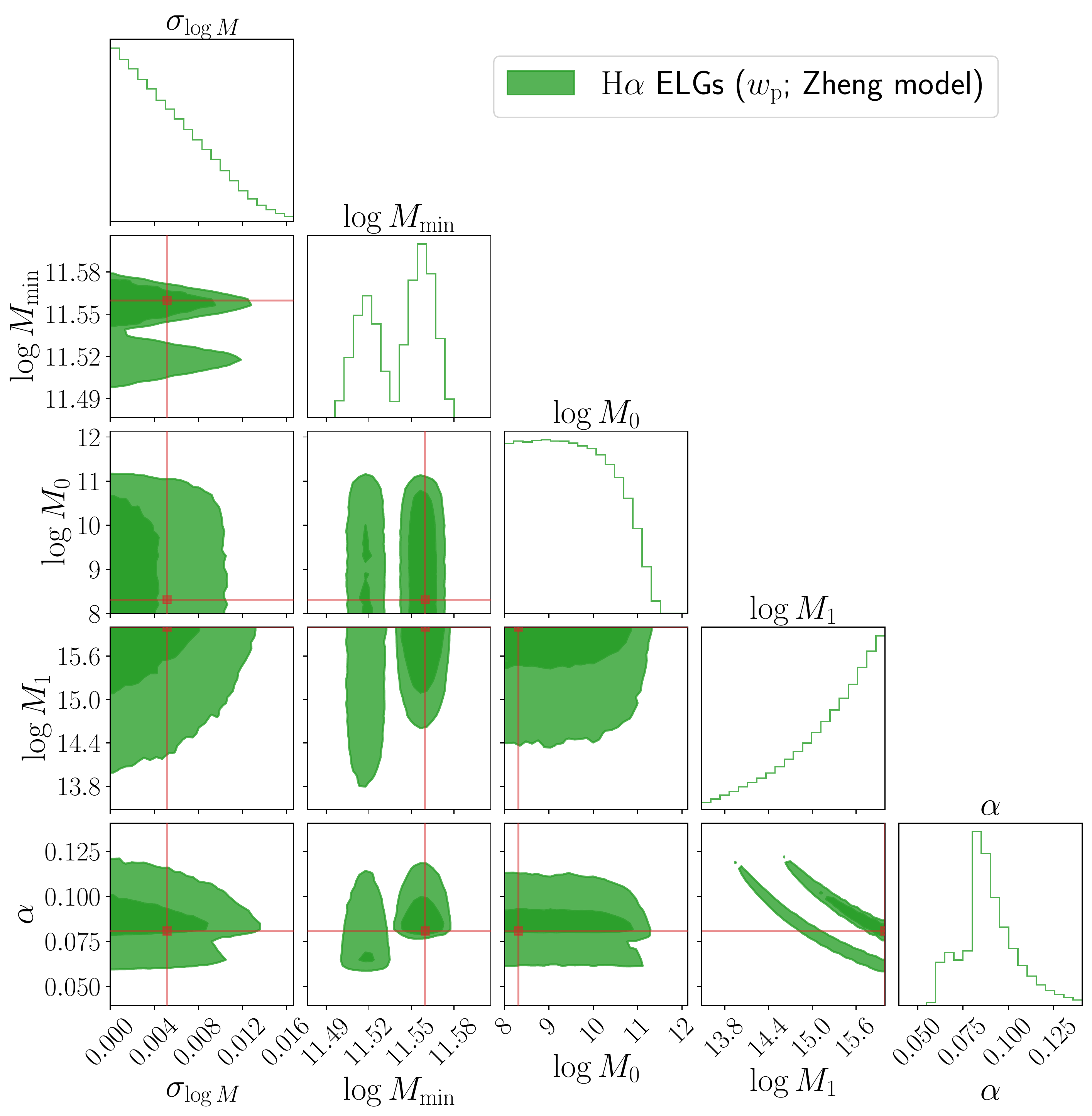}
  \caption{HOD parameter constraints with the Zheng HOD model from the projected correlation
  functions of \HA ELGs.
  The thick (thin) green regions correspond to $1\sigma$ ($2\sigma$) confidence level.
  The red lines indicate best-fit values. The unit of mass parameters
  ($M_\mathrm{min}$, $M_0$, $M_1$) is $\hMsun$.}
  \label{fig:Zheng_HA_wp_triangle}
\end{figure}

\begin{figure}
  \includegraphics[width=\columnwidth]{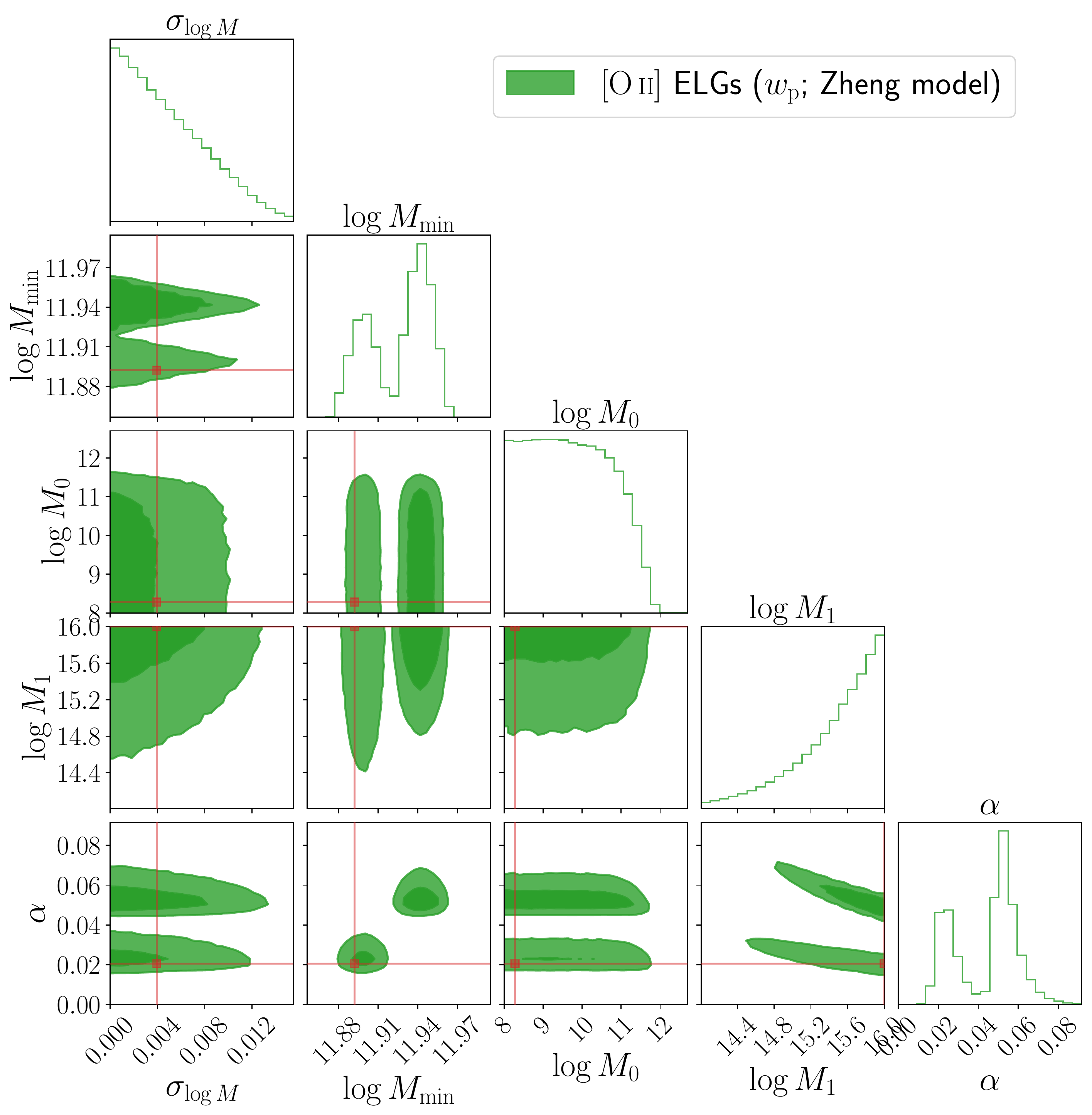}
  \caption{Same as Figure~\ref{fig:Zheng_HA_wp_triangle}
  but for \OII ELGs with the Zheng HOD model.}
  \label{fig:Zheng_OII_wp_triangle}
\end{figure}

\begin{figure*}
  \includegraphics[width=\textwidth]{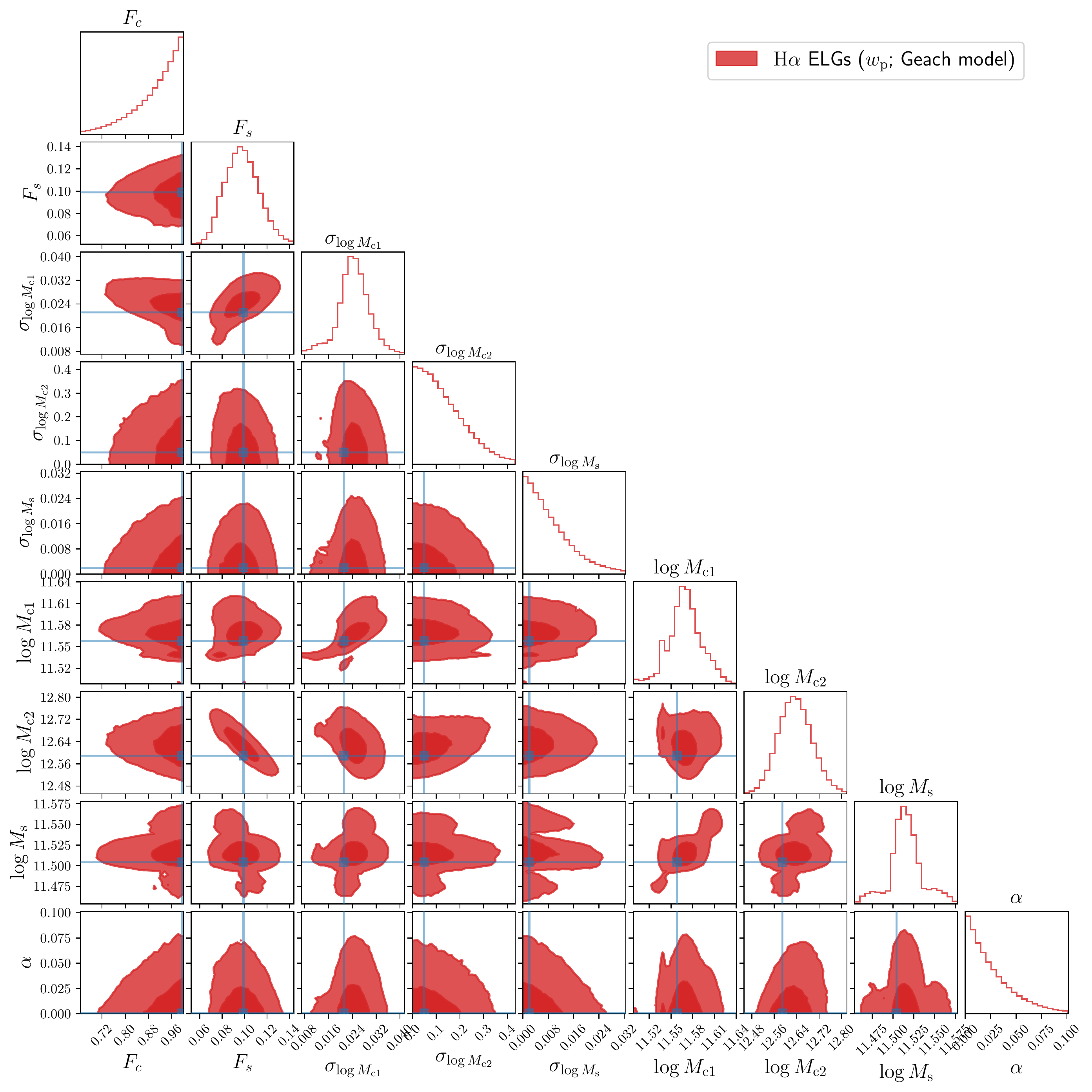}
  \caption{Same as Figure~\ref{fig:Zheng_HA_wp_triangle}
  but for \HA ELGs with the Geach HOD model.
  The unit of mass parameters
  ($M_\mathrm{c1}$, $M_\mathrm{c2}$, $M_\mathrm{s}$) is $\hMsun$.}
  \label{fig:Geach_HA_wp_triangle}
\end{figure*}

\begin{figure*}
  \includegraphics[width=\textwidth]{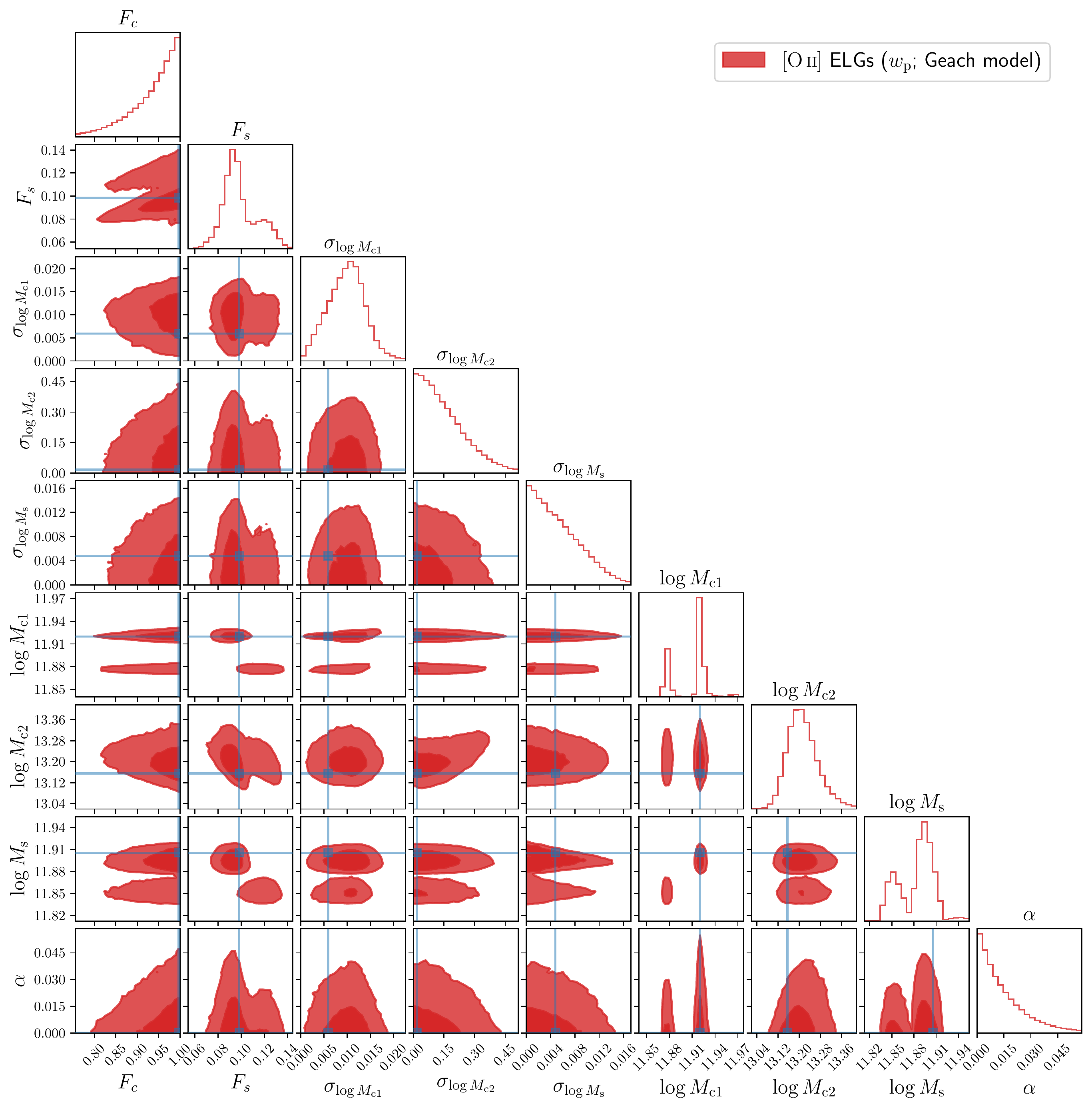}
  \caption{Same as Figure~\ref{fig:Zheng_HA_wp_triangle}
  but for \OII ELGs with the Geach HOD model.
  The unit of mass parameters
  ($M_\mathrm{c1}$, $M_\mathrm{c2}$, $M_\mathrm{s}$) is $\hMsun$.}
  \label{fig:Geach_OII_wp_triangle}
\end{figure*}

\begin{figure}
  \includegraphics[width=\columnwidth]{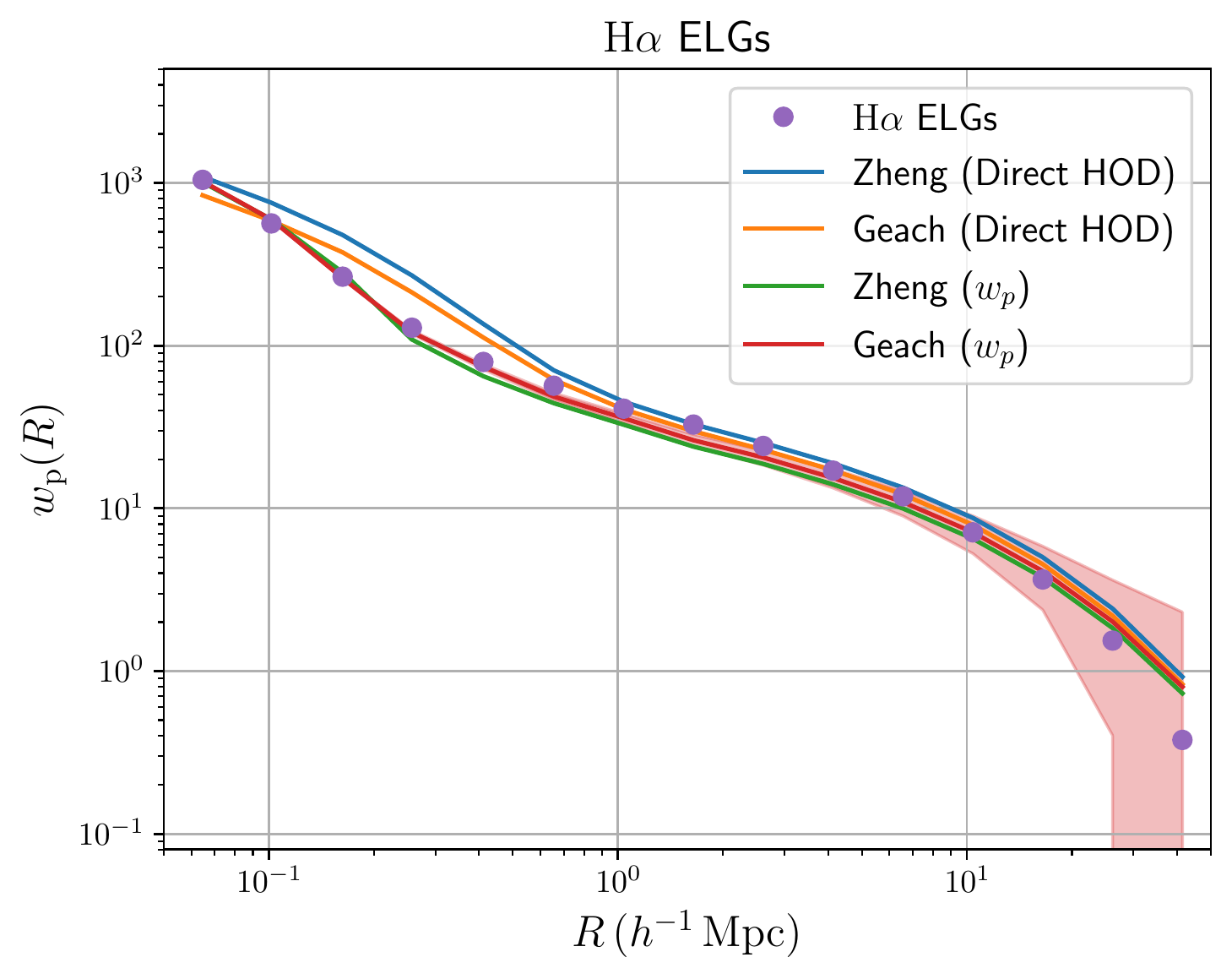}
  \includegraphics[width=\columnwidth]{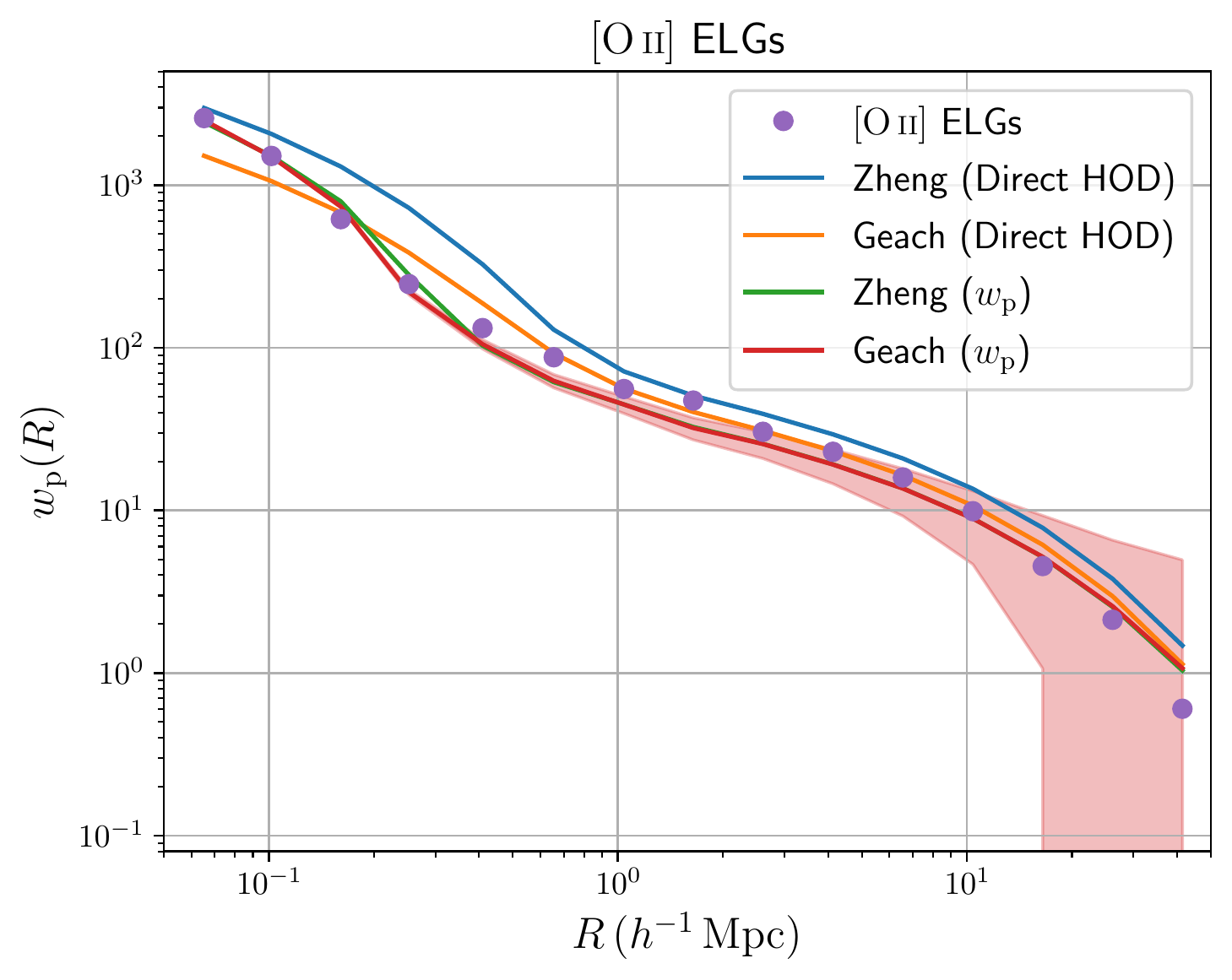}
  \caption{The projected correlation functions of \HA ELGs (upper panel)
  and \OII ELGs (lower panel).
  The results of the best-fitting direct HOD
  and best-fitting projected correlation functions
  with the Zheng and Geach models are also shown as solid lines.
  The shaded regions correspond to standard deviation derived from
  the covariance matrix with best-fit HOD parameters.
  For visual reason, we only show the error of the result of best-fitting
  projected correlation functions with the Geach model.}
  \label{fig:wp}
\end{figure}

\begin{figure*}
	\includegraphics[width=\textwidth]{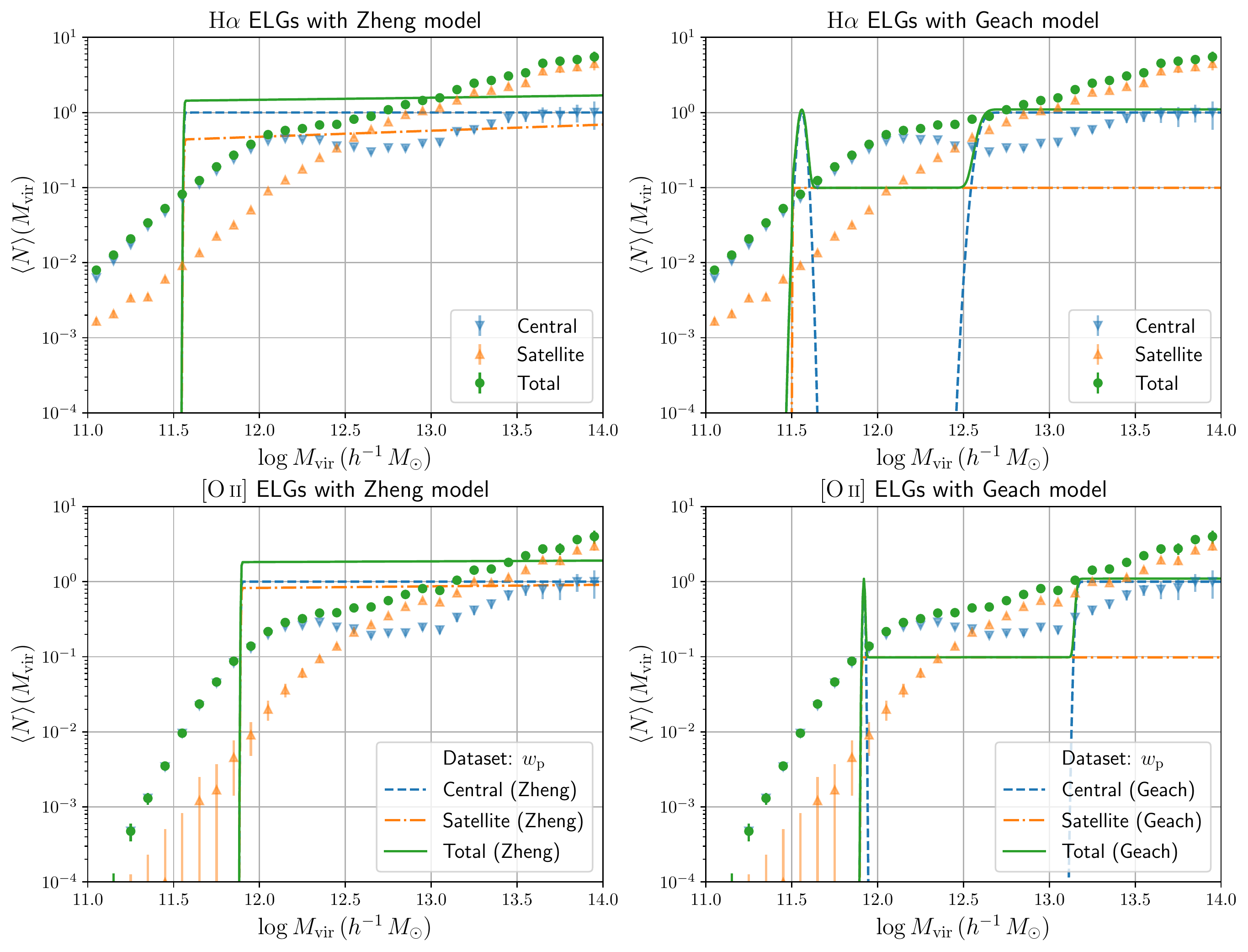}
  \caption{Best-fit HODs inferred from the projected correlation functions
  of \HA ELGs (upper panels) and \OII ELGs (lower panels)
  with the Zheng (left panels) and Geach (right panels) models.
  For comparison, the HODs measured from simulations, i.e.,
  direct HOD, are also shown as points.}
  \label{fig:bestfit_HOD_from_wp}
\end{figure*}

Figures~\ref{fig:Zheng_HA_wp_triangle}, \ref{fig:Zheng_OII_wp_triangle},
\ref{fig:Geach_HA_wp_triangle}, and \ref{fig:Geach_OII_wp_triangle}
show HOD parameter constraints with the Zheng and Geach models
from the projected correlation functions of \HA and \OII ELGs.
Though there are only 15 data points of the projected correlation function
and 1 data point of galaxy number density, the HOD parameters are well constrained.
On the other hand, several parameters, e.g., $M_\mathrm{min}$ for the Zheng model,
exhibit multi-modal features, which were not seen in the direct HOD case.
Figure~\ref{fig:wp} shows the best-fit projected correlation functions.
With the help of many HOD parameters (5 for the Zheng model and 9 for the Geach model),
both of the models can explain well the measured projected correlation function
on all scales.
On the other hand, the model predictions with best-fit parameters from direct HOD
can reproduce the projected correlation functions only for scales $\gtrsim 1 \, \hMpc$.
The Geach model slightly outperforms the Zheng model as expected,
since it fits the direct HOD better.
The best-fitting Geach model works at the reasonable level
though the non-linear effects cannot fully be captured.
The results highlight that in the case of fitting of the projected correlation function,
the advantage of the Geach model over the Zheng model is not so clear as in direct HOD inference
and the performance of both of the HOD models looks quite similar.
This fact is also indicated in the observations of \citet{Okumura2021}.
Figure~\ref{fig:bestfit_HOD_from_wp} illustrates the HODs inferred
from the projected correlation functions.
Clearly, the best-fit HODs deviate from the measured HODs.
For the Zheng model, central HOD is a sharp step-like function
and the slope of the satellite HOD is close to zero.
For the Geach model, the similar feature is observed and
for central HODs, steep log-normal contribution is preferred.
Thus, we can conclude that both of the Zheng and Geach models can explain
the observed projected correlation functions but
both of the models fail to reproduce the underlying HODs.
In order to correctly reproduce the HODs,
prior knowledge from simulations about the shape of HODs is essential.

\subsection{Summary of best-fit parameters and goodness of fit}
We have carried out parameter inference analyses with direct HOD and projected correlation functions.
Here, we summarise the best-fit HOD parameters and derived parameters from the analyses.

\begin{figure*}
  \includegraphics[width=\textwidth]{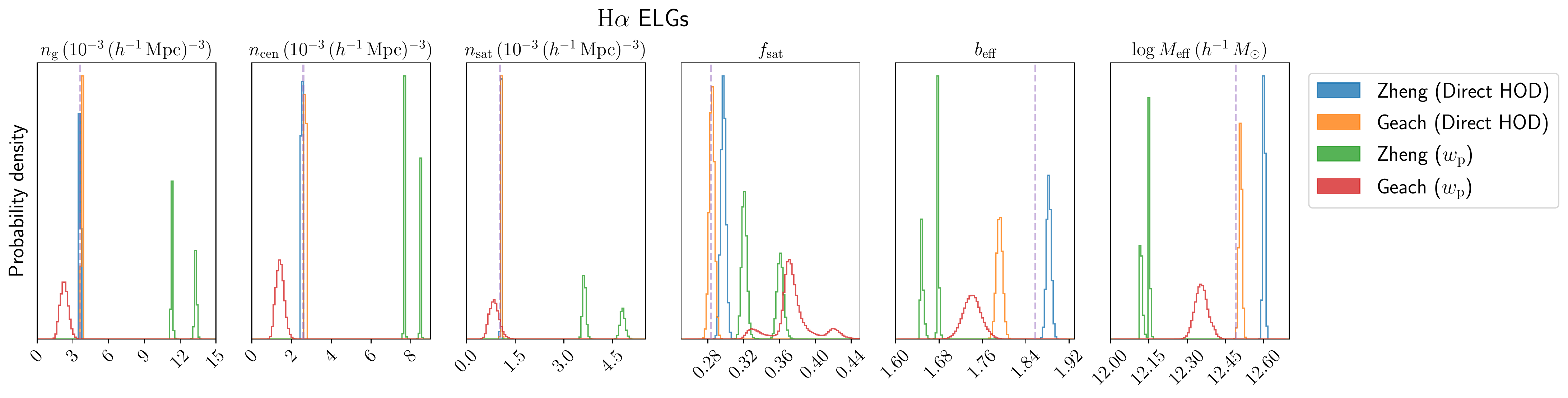}
  \includegraphics[width=\textwidth]{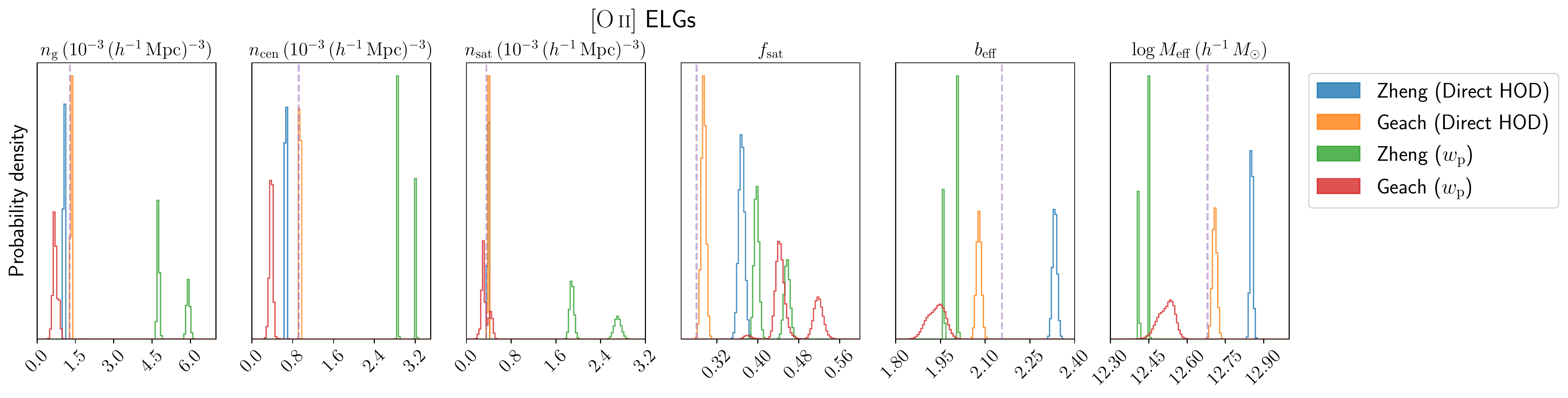}
  \caption{The one-dimensional posterior distributions of derived parameters:
  the galaxy number density $n_\mathrm{g}$,
  the central galaxy number density $n_\mathrm{cen}$,
  the satellite galaxy number density $n_\mathrm{sat}$,
  the satellite fraction $f_\mathrm{sat}$,
  the effective bias $b_\mathrm{eff}$,
  and the effective mass $M_\mathrm{eff}$.
  The purple dashed lines show the values measured directly from simulations.}
  \label{fig:histogram_derived}
\end{figure*}

Tables~\ref{tab:bestfit_Zheng_HOD_params} and \ref{tab:bestfit_Geach_HOD_params}
summarise the best-fit HOD parameters from direct HOD and projected correlation functions.
Note that the survey volume, which is matched to simulation volume $8.6 \times 10^6 \, (\hMpc)^3$,
is not so large compared with current and forthcoming spectroscopic surveys,
and thus, the covariance is larger than the practical observations.
Therefore, the current analysis
leads to the conservative constraints on the parameters.
Table~\ref{tab:derived_params} compares the derived parameters with best-fit HOD models
with the values measured from simulations. Figure~\ref{fig:histogram_derived}
shows one-dimensional histograms for derived parameters.
The galaxy number densities of central and satellite and satellite fractions
are computed directly from the mock ELG catalogues.
The effective mass $M_\mathrm{eff}$ is calculated as the mean halo mass weighted with
number of hosted ELGs. The effective bias $b_\mathrm{eff}$ can be computed in a similar manner
with the halo bias fitting formula \citep{Tinker2010}.
The general trend is that for direct HOD,
the estimates with the Geach model are closer to simulations than the Zheng model.
However, in the case of projected correlation functions,
both of the models cannot reproduce the simulations values
as these models fail to reconstruct the HOD as well.
It should be noted that the derived parameters with the best-fitting models
for projected correlation functions considerably deviate from the values measured
directly from simulations.
In particular, though the best-fitting Zheng model 
fits the projected correlation function very well,
the number density is completely inconsistent
with the true value directly measured from simulations.
Therefore, even best-fitting models lead to biased estimates of
derived parameters.

Compared with forecasts for upcoming surveys,
the number density of our mock ELGs is higher,
e.g., the forecast of the number density of \OII ELGs with PFS
based on COSMOS Mock Catalogue \citep{Jouvel2009} is
$n_\mathrm{g} = 5.5 \times 10^{-4} \, (\hMpc)^{-3}$ at the redshift $1.4 < z < 1.6$
with the flux limit $f > 5 \times 10^{-17} \, \mathrm{erg} \, \mathrm{s}^{-1} \, \mathrm{cm}^{-2}$
whereas for our mock \OII ELGs, the number density is $n_\mathrm{g} = 1.3 \times 10^{-3} (\hMpc)^{-3}$
with the flux limit $f > 6.7 \times 10^{-17} \, \mathrm{erg} \, \mathrm{s}^{-1} \, \mathrm{cm}^{-2}$
derived from the luminosity threshold $L > 10^{42} \, \mathrm{erg} \, \mathrm{s}^{-1}$
at the redshift $z = 1.5$.
The main reason for this overestimate is that no magnitude cut is applied
for our mock ELG catalogues, but in reality,
multiple magnitude or colour cuts are applied in target selection.
Since our simulation pipeline also computes broad band magnitudes,
it is possible to apply such magnitude cuts to mock ELGs and simulate target selection
\citep{Hadzhiyska2021}.
Though it is beyond the scope of this paper, the realistic target selection
will be addressed in a future study.
In terms of the satellite fraction, the satellite fractions for both of \HA and \OII ELGs
are $f_\mathrm{sat} \simeq 28\%$, i.e., the majority of ELGs is central
and is likely to be hosted by isolated halos.
\citet{Yuan2022} found that the satellite fraction
is $33\%$ for the fiducial HOD model based on hydrodynamical simulations
and \citet{Avila2020} estimated the satellite fraction as $22\%$ for the best-fit HOD model
based on mock simulations designed for eBOSS.
Though there are some scatters depending on assumed HOD models,
the satellite fractions estimated from our mock ELGs are consistent with previous studies.

\begin{table*}
  \caption{Inferred HOD parameters for the Zheng model.
  The central values are best-fit values and errors are standard deviations.
  Note that the posterior distribution of some parameters deviates from normal distribution
  or even displays multi-model feature.
  For such distributions, the errors should be interpreted with cares.}
  \label{tab:bestfit_Zheng_HOD_params}
  \begin{tabular}{cccccc}
    \hline
    Data & $\sigma_{\log M}$ & $\log M_\mathrm{min} \, (h^{-1} \, \Msun)$
    & $\log M_0 \, (h^{-1} \, \Msun)$ & $\log M_1 \, (h^{-1} \, \Msun)$ & $\alpha$ \\
    \hline \hline
    \multicolumn{6}{c}{\HA ELGs} \\
    \hline
    Direct HOD & $0.8296 \pm 0.0058$ & $12.428 \pm 0.007$ & $8.240 \pm 0.676$ & $12.786 \pm 0.009$ & $0.653 \pm 0.008$ \\
    $w_\mathrm{p}$ & $0.0052 \pm 0.0038$ & $11.560 \pm 0.022$ & $8.314 \pm 0.884$ & $16.000 \pm 0.593$ & $0.081 \pm 0.017$ \\
    \hline
    \multicolumn{6}{c}{\OII ELGs} \\
    \hline
    Direct HOD & $0.7389 \pm 0.0048$ & $12.843 \pm 0.010$ & $11.739 \pm 0.024$ & $12.990 \pm 0.028$ & $0.384 \pm 0.017$ \\
    $w_\mathrm{p}$ & $0.0039 \pm 0.0036$ & $11.892 \pm 0.023$ & $8.277 \pm 1.004$ & $15.999 \pm 0.480$ & $0.021 \pm 0.016$ \\
    \hline
  \end{tabular}
\end{table*}

\begin{table*}
  \caption{Inferred HOD parameters for the Geach model. The central values are best-fit
  values and errors are standard deviations. Since the caveat about errors also holds here,
  see the caption of Table~\ref{tab:bestfit_Zheng_HOD_params}.}
  \label{tab:bestfit_Geach_HOD_params}
  \begin{tabular}{cccccc}
    \hline
    Data & $F_\mathrm{c}$ & $F_\mathrm{s}$ & $\sigma_{\log M_\mathrm{c1}}$
    & $\sigma_{\log M_\mathrm{c2}}$ & $\sigma_{\log M_\mathrm{s}}$ \\
    \hline \hline
    \multicolumn{6}{c}{\HA ELGs} \\
    \hline
    Direct HOD & $0.319 \pm 0.007$ & $4.479 \pm 0.766$ & $0.3054 \pm 0.0057$ & $1.2598 \pm 0.0216$ & $0.9258 \pm 0.0219$ \\
    $w_\mathrm{p}$ & $0.995 \pm 0.085$ & $0.099 \pm 0.015$ & $0.0211 \pm 0.0057$ & $0.0498 \pm 0.1001$ & $0.0020 \pm 0.0078$ \\
    \hline
    \multicolumn{6}{c}{\OII ELGs} \\
    \hline
    Direct HOD & $0.261 \pm 0.009$ & $0.194 \pm 0.032$ & $0.2856 \pm 0.0049$ & $0.7271 \pm 0.0633$ & $0.4195 \pm 0.0184$ \\
    $w_\mathrm{p}$ & $0.997 \pm 0.059$ & $0.098 \pm 0.015$ & $0.0059 \pm 0.0042$ & $0.0163 \pm 0.1204$ & $0.0048 \pm 0.0040$ \\
    \hline
  \end{tabular}

  \begin{tabular}{ccccc}
    \hline
     & $\log M_\mathrm{c1} \, (h^{-1} \, \Msun)$
     & $\log M_\mathrm{c2} \, (h^{-1} \, \Msun)$
     & $\log M_\mathrm{s} \, (h^{-1} \, \Msun)$
     & $\alpha$ \\
    \hline \hline
    \multicolumn{5}{c}{\HA ELGs} \\
    \hline
    Direct HOD & $12.160 \pm 0.008$ & $13.245 \pm 0.037$ & $13.382 \pm 0.136$ & $0.034 \pm 0.083$ \\
    $w_\mathrm{p}$ & $11.558 \pm 0.024$ & $12.589 \pm 0.061$ & $11.504 \pm 0.021$ & $0.001 \pm 0.025$ \\
    \hline
    \multicolumn{5}{c}{\OII ELGs} \\
    \hline
    Direct HOD & $12.288 \pm 0.012$ & $13.356 \pm 0.033$ & $12.346 \pm 0.038$ & $0.722 \pm 0.047$ \\
    $w_\mathrm{p}$ & $11.920 \pm 0.023$ & $13.155 \pm 0.066$ & $11.906 \pm 0.024$ & $0.000 \pm 0.014$ \\
    \hline
  \end{tabular}
\end{table*}

\begin{table*}
  \caption{Derived parameters inferred from the direct HOD
  and the projected correlation functions
  with the Zheng and Geach models.
  The central values are best-fit values and errors are standard deviations.
  Since the caveat about errors also holds here,
  see the caption of Table~\ref{tab:bestfit_Zheng_HOD_params}.}
  \label{tab:derived_params}
  \begin{tabular}{ccccccc}
    \hline
    & $n_\mathrm{g}$ & $n_\mathrm{cen}$ & $n_\mathrm{sat}$
    & $f_\mathrm{sat}$ & $b_\mathrm{eff}$ & $\log M_\mathrm{eff}$ \\
    Unit & $10^{-3} \, (\hMpc)^{-3}$ & $10^{-3} \, (\hMpc)^{-3}$ &
    $10^{-3} \, (\hMpc)^{-3}$ & --- & --- & $h^{-1} \, \Msun$ \\
    \hline \hline
    \multicolumn{7}{c}{\HA ELGs} \\
    \hline
    Simulation & $3.627$ & $2.599$ & $1.028$ & $0.283$ & $1.858$ & $12.490$ \\
    Zheng (direct HOD) & $3.590 \pm 0.022$ & $2.521 \pm 0.019$ & $1.069 \pm 0.012$ & $0.298 \pm 0.003$ & $1.883 \pm 0.004$ & $12.601 \pm 0.004$ \\
    Geach (direct HOD) & $3.773 \pm 0.027$ & $2.698 \pm 0.024$ & $1.074 \pm 0.012$ & $0.285 \pm 0.003$ & $1.791 \pm 0.006$ & $12.508 \pm 0.006$ \\
    Zheng ($w_\mathrm{p}$) & $11.284 \pm 0.986$ & $7.678 \pm 0.404$ & $3.606 \pm 0.584$ & $0.320 \pm 0.020$ & $1.677 \pm 0.015$ & $12.150 \pm 0.016$ \\
    Geach ($w_\mathrm{p}$) & $2.272 \pm 0.341$ & $1.430 \pm 0.205$ & $0.842 \pm 0.152$ & $0.371 \pm 0.023$ & $1.755 \pm 0.017$ & $12.369 \pm 0.026$ \\
    \hline
    \multicolumn{7}{c}{\OII ELGs} \\
    \hline
    Simulation & $1.277$ & $0.919$ & $0.358$ & $0.280$ & $2.157$ & $12.680$ \\
    Zheng (direct HOD) & $1.055 \pm 0.013$ & $0.666 \pm 0.011$ & $0.389 \pm 0.007$ & $0.369 \pm 0.006$ & $2.333 \pm 0.007$ & $12.852 \pm 0.005$ \\
    Geach (direct HOD) & $1.338 \pm 0.013$ & $0.945 \pm 0.011$ & $0.393 \pm 0.007$ & $0.294 \pm 0.004$ & $2.077 \pm 0.008$ & $12.706 \pm 0.009$ \\
    Zheng ($w_\mathrm{p}$) & $5.896 \pm 0.572$ & $3.210 \pm 0.172$ & $2.687 \pm 0.401$ & $0.456 \pm 0.029$ & $1.960 \pm 0.023$ & $12.409 \pm 0.019$ \\
    Geach ($w_\mathrm{p}$) & $0.723 \pm 0.093$ & $0.407 \pm 0.033$ & $0.316 \pm 0.068$ & $0.437 \pm 0.038$ & $1.955 \pm 0.029$ & $12.539 \pm 0.032$ \\
    \hline
  \end{tabular}
\end{table*}

To discuss the goodness of fit, Tables~\ref{tab:chi_squares_HOD} and
\ref{tab:chi_squares_wp} summarise the chi-squares with best-fit HOD parameters
for inferences from direct HOD and projected correlation functions.
For direct HOD cases, the chi-square of the Geach model
is much better than the Zheng model as is clear from visual impression.
For projected correlation function cases, the fitting with the Geach model
is slightly better than the Zheng model but the advantage of the Geach model
is not manifest. Due to the large number of HOD parameters,
both of models can fit the projected correlation functions,
which in fact exhibit the monotonic shape for our fitting range.
If data points at smaller scales, where the transition from two-halo term to
one-halo term occurs, are included,
the Zheng and Geach models can be distinguishable more clearly.

\begin{table}
  \caption{The chi-squares with best-fit HOD parameters from the direct HOD.
  The number of data points is $N_\mathrm{data} = 30+30+1 = 61$,
  the number of parameters is $N_\mathrm{params} = 5$ for the Zheng model
  and $N_\mathrm{params} = 9$ for the Geach model, and the degree of freedom
  is $N_\mathrm{dof} = N_\mathrm{data} - N_\mathrm{params} = 56$
  for the Zheng model and $N_\mathrm{dof} = 52$ for the Geach model.}
  \label{tab:chi_squares_HOD}
  \begin{tabular}{cccccc}
    \hline
    & $\chi^2_\mathrm{cen}$ & $\chi^2_\mathrm{sat}$ & $\chi^2_{n_\mathrm{g}}$
    & $\chi^2_\mathrm{total}$ & $\chi^2_\mathrm{total} / N_\mathrm{dof}$\\
    \hline \hline
    \multicolumn{6}{c}{\HA ELGs} \\
    \hline
    Zheng & $3035.055$ & $155.406$ & $0.004$ & $3190.464$ & $56.973$ \\
    Geach & $88.756$ & $112.769$ & $0.055$ & $201.580$ & $3.877$ \\
    \hline
    \multicolumn{6}{c}{\OII ELGs} \\
    \hline
    Zheng & $3541.013$ & $62.061$ & $0.913$ & $3603.987$ & $64.357$ \\
    Geach & $16.428$ & $22.932$ & $0.054$ & $39.414$ & $0.758$ \\
    \hline
  \end{tabular}
\end{table}

\begin{table}
  \caption{The chi-squares with best-fit HOD parameters
  from the projected correlation functions.
  The number of data points is $N_\mathrm{data} = 15+1 = 16$,
  the number of parameters is $N_\mathrm{params} = 5$ for the Zheng model
  and $N_\mathrm{params} = 9$ for the Geach model, and the degree of freedom
  is $N_\mathrm{dof} = N_\mathrm{data} - N_\mathrm{params} = 11$
  for the Zheng model and $N_\mathrm{dof} = 7$ for the Geach model.}
  \label{tab:chi_squares_wp}
  \begin{tabular}{cccccc}
    \hline
    & $\chi^2_{w_\mathrm{p}}$ & $\log \det \mathrm{Cov}$
    & $\chi^2_{n_\mathrm{g}}$ & $\chi^2_\mathrm{total}$
    & $\chi^2_\mathrm{total} / N_\mathrm{dof}$\\
    \hline \hline
    \multicolumn{6}{c}{\HA ELGs} \\
    \hline
    Zheng & $258.802$ & $22.983$ & $45.328$ & $327.113$ & $29.738$ \\
    Geach & $116.691$ & $23.442$ & $7.693$ & $147.826$ & $21.118$ \\
    \hline
    \multicolumn{6}{c}{\OII ELGs} \\
    \hline
    Zheng & $317.360$ & $55.102$ & $58.572$ & $431.034$ & $39.185$ \\
    Geach & $168.730$ & $55.001$ & $8.100$ & $231.831$ & $33.119$ \\
    \hline
  \end{tabular}
\end{table}

\section{Conclusions}
\label{sec:conclusions}
Upcoming spectroscopic redshift surveys aim to
measure the statistics of the galaxy distribution and constrain cosmological models
at the unprecedented accuracy.
In such upcoming surveys ELGs become the main target
as a tracer of the large-scale matter distribution because they probe the deeper universe,
$z\gtrsim 1$, where star formation is active.
The formation and evolution processes of ELGs are, however,
quite different from those of LRGs
which have been employed in the past galaxy surveys. 
While LRGs are old and quenched galaxies, 
ELGs are blue and star-forming galaxies and thus
are undergoing infall towards massive halos along the filamentary structures.
This coherent dynamics has a considerable effect on modelling of galaxy clustering statistics.
Therefore, it is critical and timely to address the galaxy-halo connection for ELGs.

In this study, we leverage the hydrodynamical simulations,
where galaxy formation physics is implemented as the subgrid model.
The state-of-the-art hydrodynamical simulations can reproduce
a variety of observational results, e.g.,
stellar mass functions and stellar-to-halo mass relations.
Therefore, the simulations enable us to generate realistic mock ELG catalogues.
Among the latest galaxy formation simulations, we have employed IllustrisTNG simulations,
which cover a large cosmological volume $(205 \, \hMpc)^3$ in comoving scale.
With this simulation suite, we generate mock ELG catalogues
of \HA and \OII ELGs, which are targets of PFS and \textit{Euclid} surveys,
respectively, by post-processing the particle outputs and halo catalogues
with stellar population synthesis code \texttt{P\'EGASE-3}.
Since IllustrisTNG simulations have information about
the matter distributions and detailed halo catalogues,
we can scrutinise the ELG-halo connections,
which help us improve theoretical modelling of ELG clustering including the non-linear regime.

First we have focused on the basic properties of mock ELGs to validate
our mock catalogues.
Our mock ELG catalogues successfully reproduced
the luminosity functions and SFR-line luminosity relations
estimated from observations once the dust attenuation effect was considered.
Next, we have measured the HODs, i.e., mean occupation number of ELGs
as a function of host halo mass.
It is known that the central ELGs are composed of two distinct components:
infalling disk galaxies and bulge-dominated spheroidal galaxies.
In particular, the former induce the peak in HODs of central HODs,
which is well described as the log-normal distribution.
This feature has been confirmed also from our mock ELG catalogues
and the Geach model explained the shape of HODs better than the conventional Zheng model.

Then, we have carried out the inference analysis of HOD parameters
using the HODs directly measured from the mock catalogues.
As expected, the Geach model can fit the measured HODs better and reproduce
the galaxy number density as well.
The Zheng model has fewer parameters and cannot describe the specific shape of central HODs.
We have also investigated the constraints of HOD parameters from
the projected correlation functions, which are widely used as cosmological statistics
of the ELG distribution.
Both of the Zheng and Geach models can reproduce the measured correlation functions
down to small scales ($\simeq 0.1 \, \hMpc$).
These models have many free parameters ($5$ and $9$ parameters for the Zheng and Geach models, respectively) and thus,
the large degree of freedom helps to fit the entire shape of the projected correlation functions.
The HODs with best-fit parameters are, however, appreciably different from the measured HODs.
It is expected
because the projected correlation functions entail only a part of information of
ELG clustering.

In this paper, as Part I of the series, we have focused only on the projected correlation function,
which is less subject to the velocity information of ELGs.
The kinematics of ELG in host halos, e.g., infall motion, strongly affects the velocity statistics,
which can be observed through redshift space distortions.
In the forthcoming papers of this series, we will present
a challenge analysis with three-dimensional ELG correlation functions in redshift space
to constrain the growth rate.
Then we will investigate the relationship between ELGs and host halos,
paying particular attention to the velocity and assembly biases of ELGs.
That will help us construct more reliable modelling of ELG clustering
and enhance the capability of theoretical models to predict
the cosmological statistics precisely.

\section*{Acknowledgements}
K.~O. is supported by JSPS Research Fellowships for Young Scientists.
This work was supported by Grant-in-Aid for JSPS Fellows Grant Number JP21J00011
and Grant-in-Aid for Early-Career Scientists Grant Number JP22K14036.
T.~O. acknowledges support from the Ministry of
Science and Technology of Taiwan under Grants No. MOST
110-2112-M-001-045- and the Career Development Award,
Academia Sinina (AS-CDA-108-M02) for the period of 2019 to 2023.
We acknowledge the IllustrisTNG team to make the simulation data publicly available.
We thank Takahiro Nishimichi, Shun Saito, Jingjing Shi, Masahiro Takada,
and PFS cosmology working group for fruitful discussions on the earlier results.
Numerical computations were carried out
on Cray XC50 at Center for Computational Astrophysics, National Astronomical Observatory of Japan
and FUJITSU Supercomputer PRIMEHPC FX1000 (Wisteria-O, Wisteria/BDEC-01)
at Information Technology Center, The University of Tokyo.

\section*{Data Availability}
The data underlying this article
will be shared on reasonable request to the corresponding author.
The simulation data of IllustrisTNG is available at \url{https://www.tng-project.org/}.



\bibliographystyle{mnras}
\bibliography{main}



\appendix

\section{Convergence with respect to resolution of hydrodynamical simulations}
\label{sec:convergence}
\begin{figure}
  \includegraphics[width=\columnwidth]{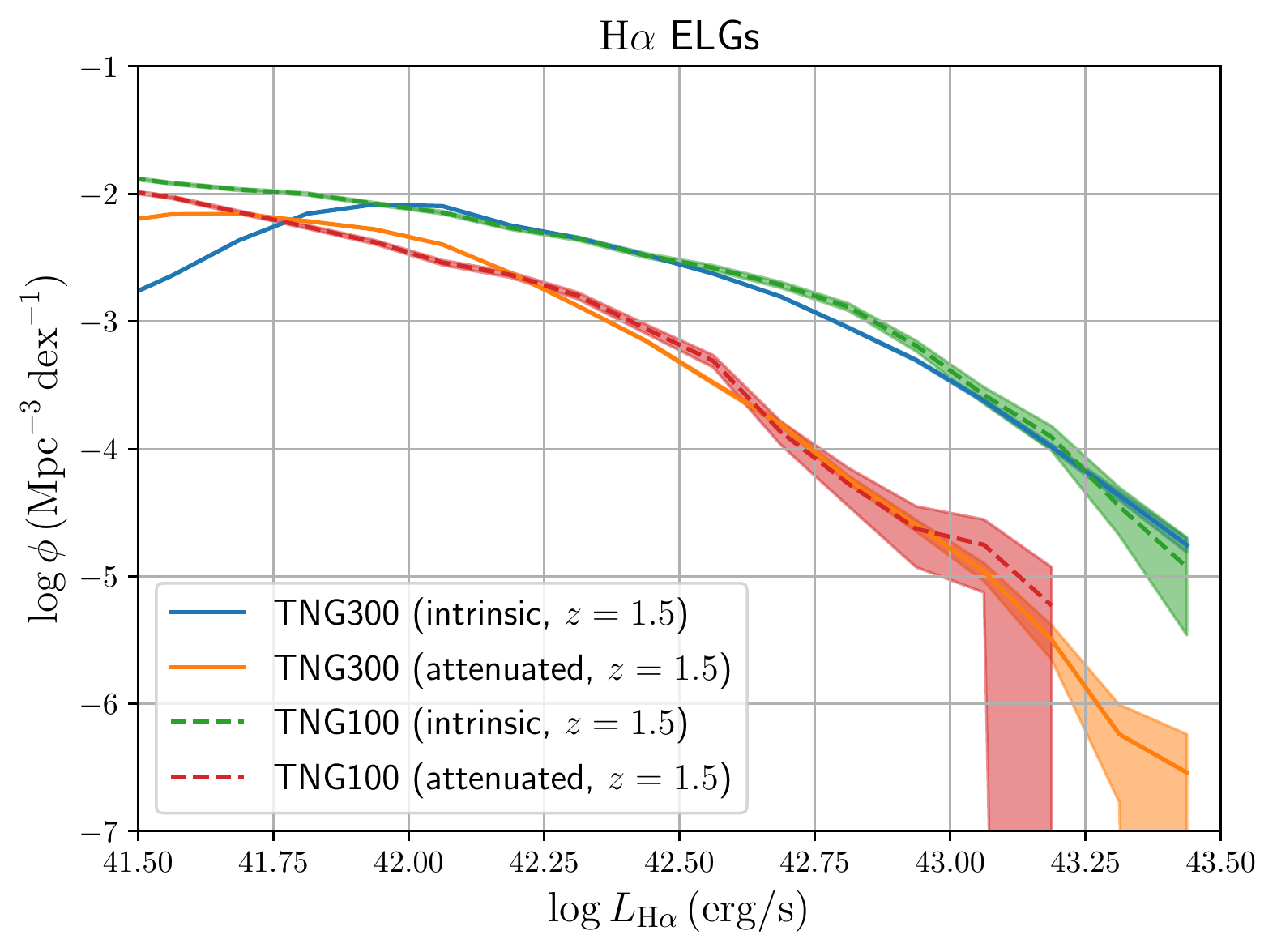}
  \includegraphics[width=\columnwidth]{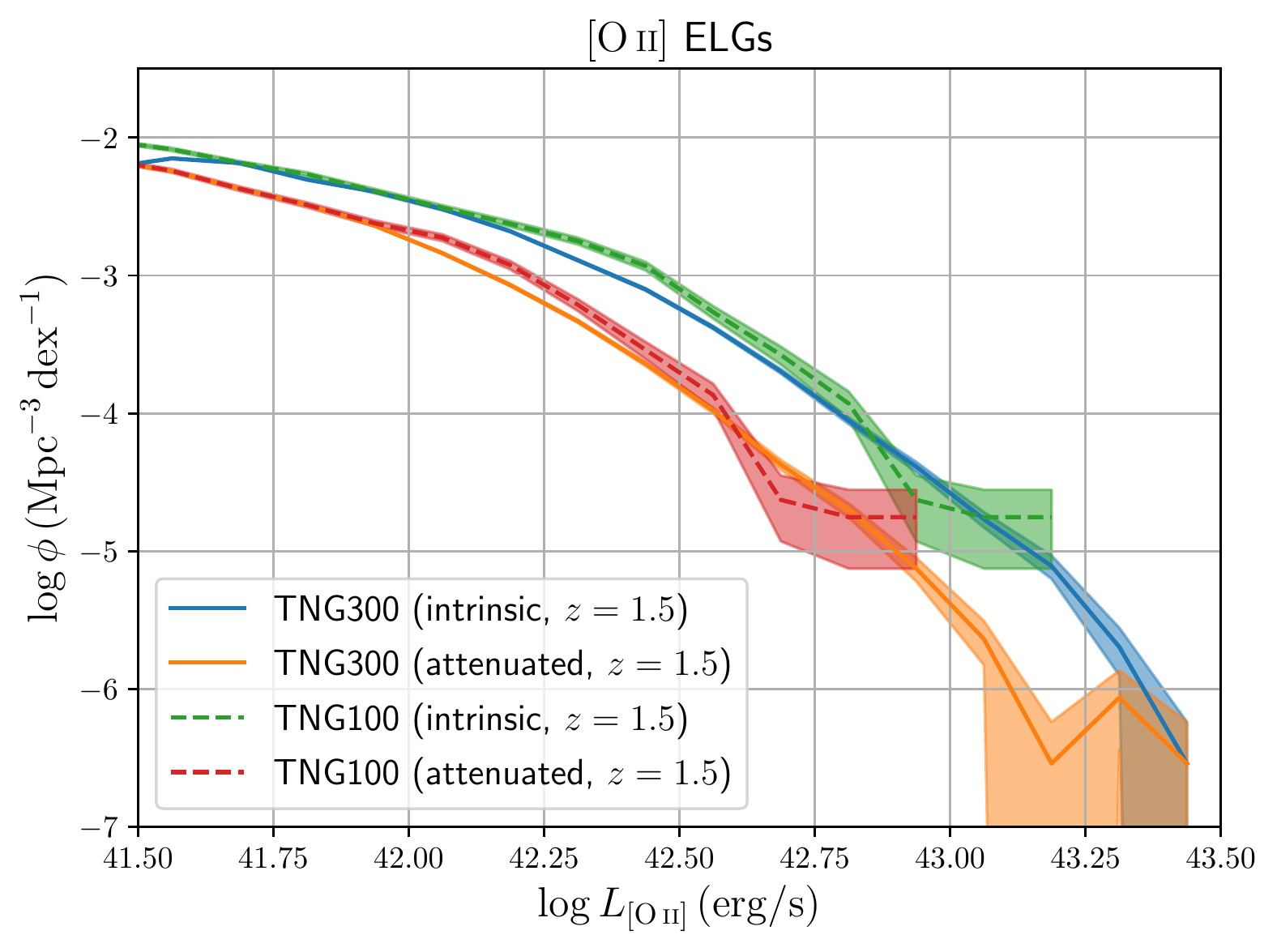}
  \caption{Luminosity functions of \HA ELGs (upper panel)
  and \OII ELGs (lower panel) with TNG300 and TNG100.
  Orange and red solid lines show the results
  with the dust attenuation effect considered but
  blue and green solid lines correspond to the results of intrinsic luminosities.
  The error bars are determined as Poisson error.}
  \label{fig:LF_res}
\end{figure}

In order to address the effects of resolution of the simulations,
we compare the results with TNG300, which is the fiducial simulation used in this work,
and the higher resolution run TNG100-1, hereafter TNG100.
For TNG100, the simulation box length on a side is $75 \, \hMpc$
and the number of dark matter particle and gas cells is $1820^3$ for each.
The resultant mass resolution is $5.1 \times 10^6 \, \hMsun$ for dark matter particle
and $9.4 \times 10^5 \, \hMsun$ for gas cell.
Figure~\ref{fig:LF_res} shows the luminosity functions of \HA and \OII ELGs
with TNG300 and TNG100.
Overall, the results from these two simulations are consistent.
Since TNG100 simulation can resolve faint ELGs,
the luminosity functions without dust attenuation keep
the power-law form down to $10^{41.5} \, \mathrm{erg} \, \mathrm{s}^{-1}$,
where the sudden decrease appears for TNG300 due to the lack of resolution.
In this work, we have applied luminosity threshold $L > 10^{42}\, \mathrm{erg} \, \mathrm{s}^{-1}$
to define the ELG samples and
we can conclude that the resolution convergence is well achieved.

\section{Effect of the scale cut for HOD parameter inference with the projected correlation function}
\label{sec:wp_cut}
\begin{figure}
  \includegraphics[width=\columnwidth]{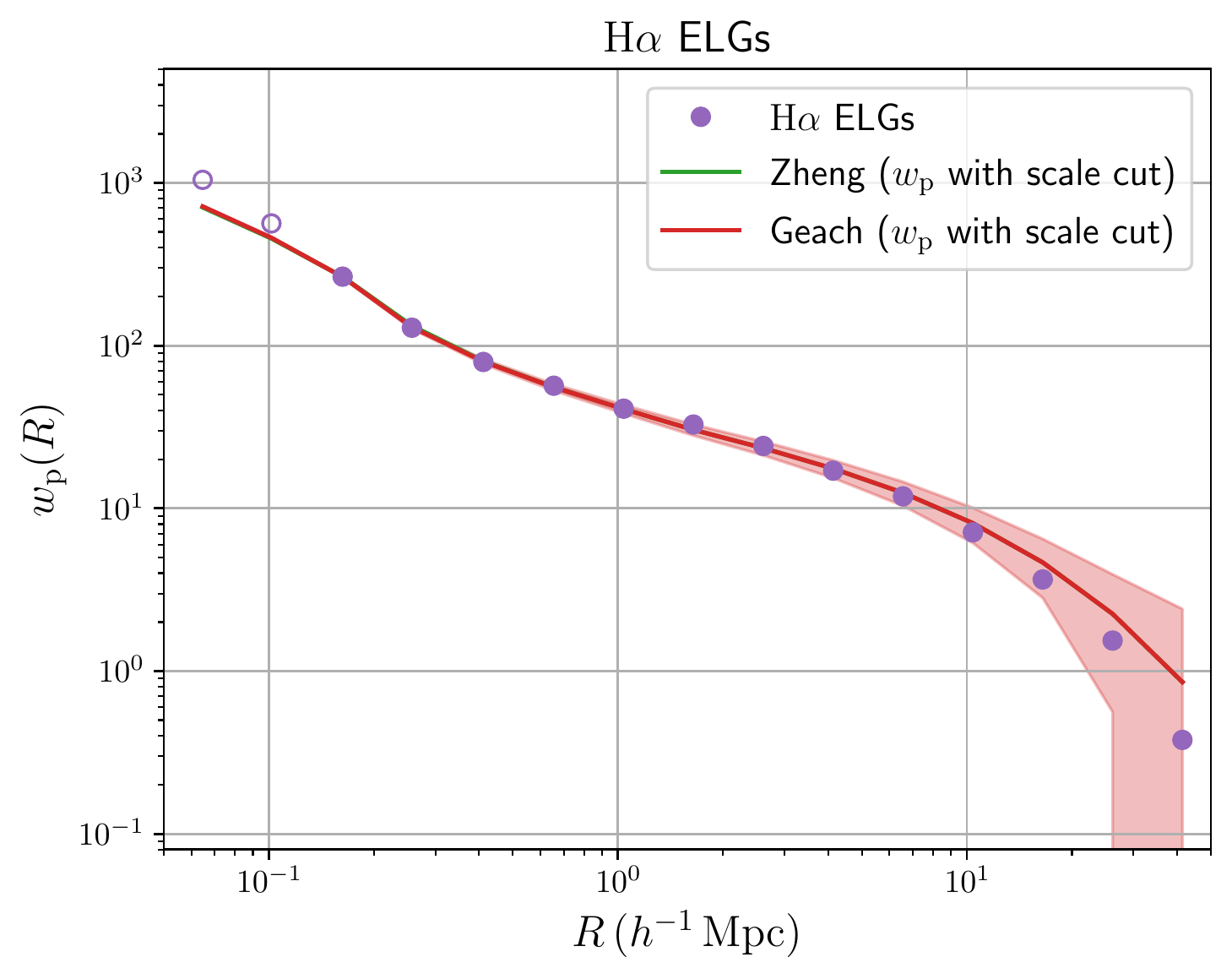}
  \includegraphics[width=\columnwidth]{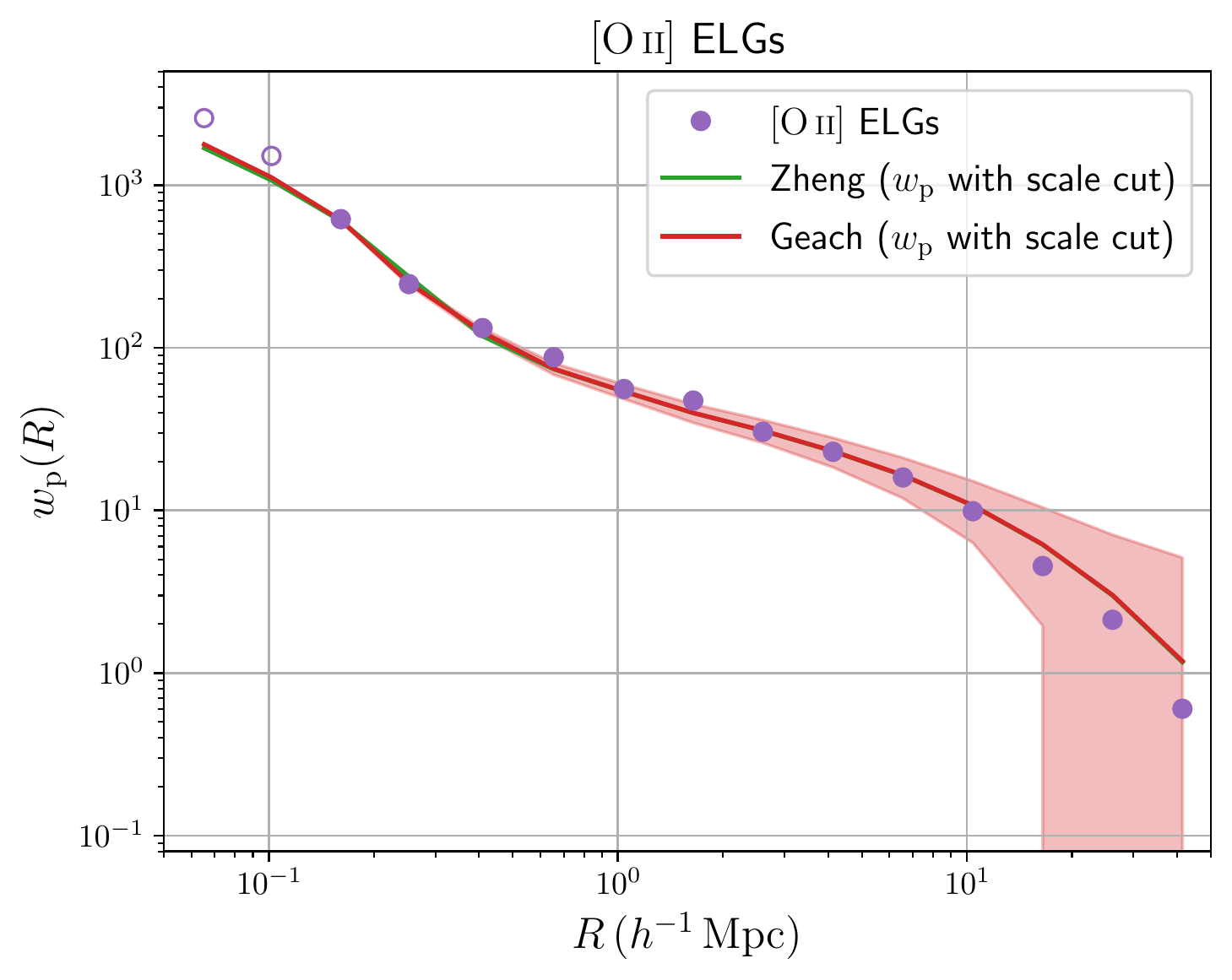}
  \caption{The projected correlation functions of \HA ELGs (upper panel)
  and \OII ELGs (lower panel) with the scale cut.
  The shaded regions correspond to standard deviation derived from
  the covariance matrix with best-fit HOD parameters.
  For a visual reason, we only show the error of the result of best-fitting
  projected correlation functions with the Geach model.
  The open symbols correspond to the data points excluded in the analysis due to the scale cut.}
  \label{fig:wp_cut}
\end{figure}

\begin{figure*}
	\includegraphics[width=\textwidth]{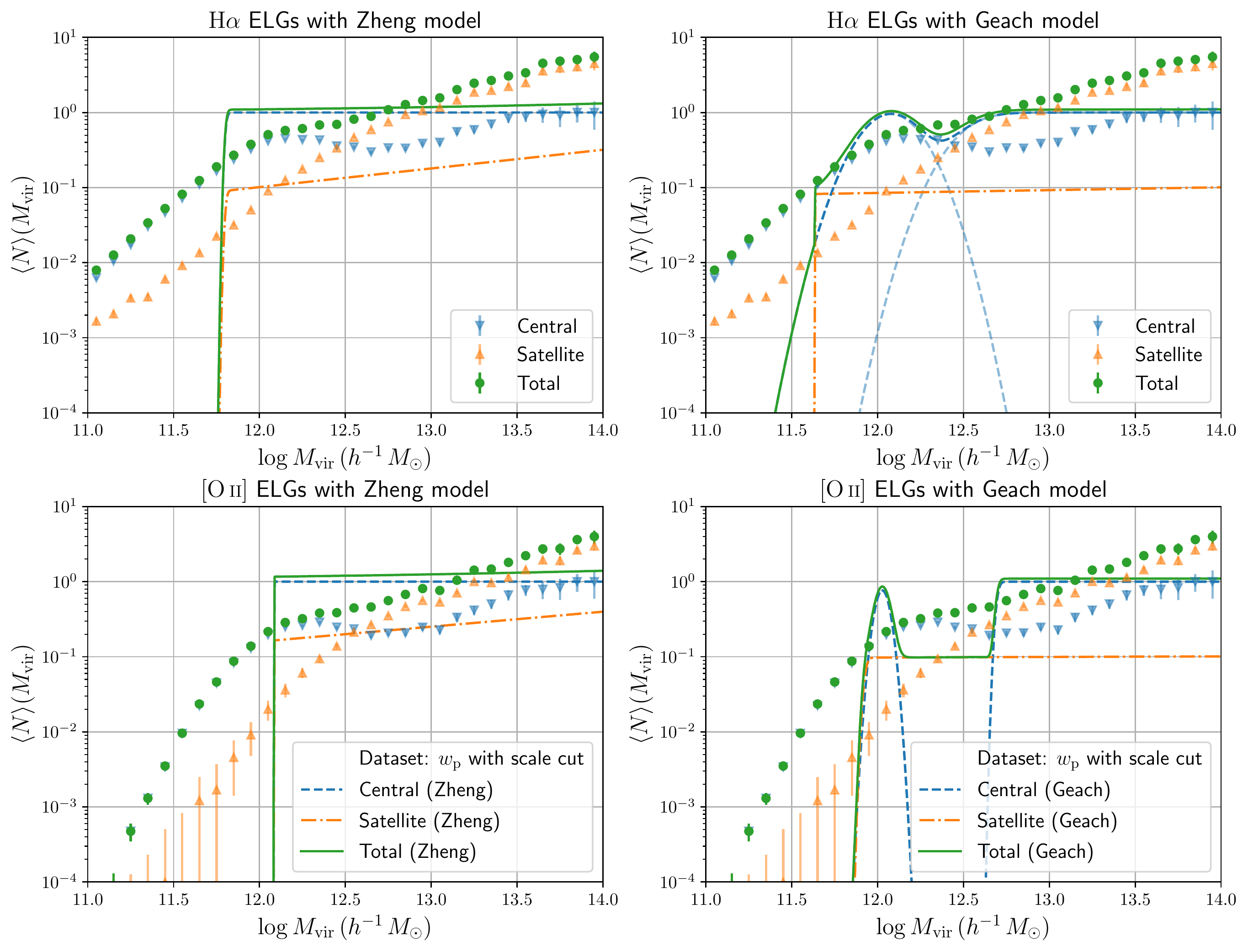}
  \caption{Best-fit HODs inferred from the projected correlation functions
  of \HA ELGs (upper panels) and \OII ELGs (lower panels) with the scale cut
  with the Zheng (left panels) and Geach (right panels) models.
  For comparison, the HODs measured from simulations, i.e.,
  direct HOD, are also shown as points.}
  \label{fig:bestfit_HOD_from_wp_cut}
\end{figure*}

While the theoretical prediction of the projected correlation function based on the halo model is less accurate at small scales, 
the error of the measurement at such scales becomes smaller. Thus, the parameter inference strongly depends on
the small-scale data points and inferred parameters can be biased due to the low accuracy of the model.
Here, we investigate how the small-scale data points affect the fitting of the projected correlation function.
In order to address this effect, we have carry out the HOD parameter inference
with the projected correlation functions with the scale cut
which excludes two data points at $R = 6.46 \times 10^{-2} \, h^{-1} \, \mathrm{Mpc}$
and $1.01 \times 10^{-1} \, h^{-1} \, \mathrm{Mpc}$. 
Since removing these data points leads to worse convergence of our MCMC analysis,
we show only the best-fit correlation functions and HODs.
Figure~\ref{fig:wp_cut} shows the best-fit projected correlation functions with the scale cut.
Clearly, the projected correlation functions can be reproduced well similarly to the case without the scale cut.
Next, Figure~\ref{fig:bestfit_HOD_from_wp_cut} shows the best-fit HODs inferred from the projected correlation functions
with the scale cut. The best-fit parameters for the Zheng and Geach models
are summarized in Tables~\ref{tab:bestfit_Zheng_HOD_params_wp_cut}
and \ref{tab:bestfit_Geach_HOD_params_wp_cut}, respectively.
The derived parameters and goodness-of-fit with the best-fit HOD parameters are respectively shown
in Tables~\ref{tab:derived_params_wp_cut}
and \ref{tab:chi_squares_wp_cut}.
The scale cut yields some improvements in reproducing true HODs and galaxy number densities directly measured in simulations.
For example, for both \HA and \OII ELGs, the peak locations of central HODs become closer
to those of measured HODs than without the scale cut.
However, the overall shape of HODs is still quite different from the measured HODs.
Thus, our conclusion that HODs cannot be fully reproduced solely with the projected correlation functions
holds even with the scale cut.

\begin{table*}
  \caption{The best-fit HOD parameters inferred with the projected correlation functions
  with the scale cut for the Zheng model.}
  \label{tab:bestfit_Zheng_HOD_params_wp_cut}
  \begin{tabular}{cccccc}
    \hline
    Data & $\sigma_{\log M}$ & $\log M_\mathrm{min} \, (h^{-1} \, \Msun)$
    & $\log M_0 \, (h^{-1} \, \Msun)$ & $\log M_1 \, (h^{-1} \, \Msun)$ & $\alpha$ \\
    \hline \hline
    \multicolumn{6}{c}{\HA ELGs} \\
    \hline
    $w_\mathrm{p}$ with scale cut & $0.0159$ & $11.801$ & $9.244$ & $15.998$ & $0.249$ \\
    \hline
    \multicolumn{6}{c}{\OII ELGs} \\
    \hline
    $w_\mathrm{p}$ with scale cut & $0.0005$ & $12.083$ & $8.244$ & $15.999$ & $0.200$ \\
    \hline
  \end{tabular}
\end{table*}

\begin{table*}
  \caption{The best-fit HOD parameters inferred with the projected correlation functions
  with the scale cut for the Geach model.}
  \label{tab:bestfit_Geach_HOD_params_wp_cut}
  \begin{tabular}{ccccccccccc}
    \hline
    Data & $F_\mathrm{c}$ & $F_\mathrm{s}$ & $\sigma_{\log M_\mathrm{c1}}$
    & $\sigma_{\log M_\mathrm{c2}}$ & $\sigma_{\log M_\mathrm{s}}$
    & $\log M_\mathrm{c1} \, (h^{-1} \, \Msun)$
    & $\log M_\mathrm{c2} \, (h^{-1} \, \Msun)$
    & $\log M_\mathrm{s} \, (h^{-1} \, \Msun)$
    & $\alpha$ \\\\
    \hline \hline
    \multicolumn{10}{c}{\HA ELGs} \\
    \hline
    $w_\mathrm{p}$ with scale cut & $0.956$ & $0.082$ & $0.1574$ & $0.2204$ & $0.0015$ & $12.077$ & $12.474$ & $11.634$ & $0.038$ \\
    \hline
    \multicolumn{10}{c}{\OII ELGs} \\
    \hline
    $w_\mathrm{p}$ with scale cut & $0.767$ & $0.098$ & $0.0403$ & $0.0238$ & $0.0258$ & $12.027$ & $12.692$ & $11.923$ & $0.008$ \\
    \hline
  \end{tabular}
\end{table*}

\begin{table*}
  \caption{The best-fit derived parameters inferred with the projected correlation functions
  with the scale cut. For comparison, the results measured from simulations are also shown.}
  \label{tab:derived_params_wp_cut}
  \begin{tabular}{ccccccc}
    \hline
    & $n_\mathrm{g}$ & $n_\mathrm{cen}$ & $n_\mathrm{sat}$
    & $f_\mathrm{sat}$ & $b_\mathrm{eff}$ & $\log M_\mathrm{eff}$ \\
    Unit & $10^{-3} \, (\hMpc)^{-3}$ & $10^{-3} \, (\hMpc)^{-3}$ &
    $10^{-3} \, (\hMpc)^{-3}$ & --- & --- & $h^{-1} \, \Msun$ \\
    \hline \hline
    \multicolumn{7}{c}{\HA ELGs} \\
    \hline
    Simulation & $3.627$ & $2.599$ & $1.028$ & $0.283$ & $1.858$ & $12.490$ \\
    Zheng ($w_\mathrm{p}$ with scale cut) & $4.571$ & $4.107$ & $0.464$ & $0.101$ & $1.878$ & $12.344$ \\
    Geach ($w_\mathrm{p}$ with scale cut) & $3.433$ & $2.903$ & $0.529$ & $0.154$ & $1.878$ & $12.361$ \\
    \hline
    \multicolumn{7}{c}{\OII ELGs} \\
    \hline
    Simulation & $1.277$ & $0.919$ & $0.358$ & $0.280$ & $2.157$ & $12.680$ \\
    Zheng ($w_\mathrm{p}$ with scale cut) & $2.374$ & $1.989$ & $0.384$ & $0.162$ & $2.153$ & $12.559$ \\
    Geach ($w_\mathrm{p}$ with scale cut) & $1.085$ & $0.790$ & $0.296$ & $0.272$ & $2.156$ & $12.631$ \\
    \hline
  \end{tabular}
\end{table*}

\begin{table}
  \caption{The chi-squares with best-fit HOD parameters
  from the projected correlation functions with the scale cut.
  The number of data points is $N_\mathrm{data} = 13+1 = 14$,
  the number of parameters is $N_\mathrm{params} = 5$ for the Zheng model
  and $N_\mathrm{params} = 9$ for the Geach model, and the degree of freedom
  is $N_\mathrm{dof} = N_\mathrm{data} - N_\mathrm{params} = 9$
  for the Zheng model and $N_\mathrm{dof} = 5$ for the Geach model.}
  \label{tab:chi_squares_wp_cut}
  \begin{tabular}{cccccc}
    \hline
    & $\chi^2_{w_\mathrm{p}}$ & $\log \det \mathrm{Cov}$
    & $\chi^2_{n_\mathrm{g}}$ & $\chi^2_\mathrm{total}$
    & $\chi^2_\mathrm{total} / N_\mathrm{dof}$\\
    \hline \hline
    \multicolumn{6}{c}{\HA ELGs} \\
    \hline
    Zheng & $5.510$ & $16.935$ & $1.883$ & $24.328$ & $2.703$ \\
    Geach & $3.448$ & $16.929$ & $0.107$ & $20.483$ & $4.097$ \\
    \hline
    \multicolumn{6}{c}{\OII ELGs} \\
    \hline
    Zheng & $27.149$ & $43.332$ & $9.621$ & $80.102$ & $8.900$ \\
    Geach & $8.955$ & $43.366$ & $0.661$ & $52.982$ & $10.596$ \\
    \hline
  \end{tabular}
\end{table}


\bsp	
\label{lastpage}
\end{document}